\documentclass[a4paper,fleqn]{cas-dc}

\usepackage{natbib}

\renewcommand{\cite}[1]{\citep{#1}}

\def\tsc#1{\csdef{#1}{\textsc{\lowercase{#1}}\xspace}}
\tsc{WGM}
\tsc{QE}

\begin{document}
\let\WriteBookmarks\relax
\def\floatpagepagefraction{1}
\def\textpagefraction{.001}

\shorttitle{Thermodynamics of Information}    

\shortauthors{J.M.R. Parrondo}  

\title [mode = title]{Thermodynamics of Information}  



%

\author[1]{Juan M.R. Parrondo}[orcid=0000-0001-8525-3709]



\ead{parrondo@ucm.es}

\ead[url]{http://seneca.fis.ucm.es/parr}


\affiliation[1]{organization={Departamento de Estructura de la Materia, F\'isica T\'ermica y Electr\'onica and  GISC, Universidad Complutense de Madrid},
            addressline={Ciudad Universitaria s/n}, 
            city={},
       citysep={}, 
            postcode={28040}, 
            state={Madrid},
            country={Spain}}




\begin{abstract}
As early as 1867, two years after the introduction of the concept of entropy by Clausius, Maxwell showed that the limitations imposed by the second law of thermodynamics   depend on the information that one possesses about the state of a physical system. A ``very observant and neat-fingered being'', later on named {\em Maxwell demon} by Kelvin, could arrange the molecules of a gas and induce a temperature or pressure gradient without performing work, in apparent contradiction to the second law. One century later, Landauer claimed that ``information is physical'', and showed that  certain  processes involving information, like overwriting a memory, need work to be completed and are unavoidably accompanied by heat dissipation. Thermodynamics of information  analyzes this bidirectional influence between thermodynamics and information processing. The seminal ideas that Landauer and Bennett devised in the 1970's have been recently reformulated in a more precise and general way by realizing that informational states are out of equilibrium  and applying new tools from non-equilibrium statistical mechanics. 
\end{abstract}


%
%
\begin{keywords}
Feedback processes \sep 
Information engines \sep 
Information flows \sep 
Landauer principle \sep 
Maxwell demon \sep 
Mutual information \sep 
Non-equilibrium free energy  \sep 
Shannon entropy \sep 
Szil\'ard engine 
\end{keywords}

\maketitle


\section{Introduction}
\label{sec:intro}


The connection between information and thermodynamics was first revealed by the celebrated  {\em Maxwell demon}, a gedanken experiment that  Maxwell devised shortly after his discovery of the distribution of velocities in an equilibrium gas. 
The velocity distribution revealed that temperature is proportional to the average kinetic energy of molecules, but their speeds cover a wide range of values. Maxwell considered two gases, a hot one $A$ and a cold one $B$, confined respectively in two containers separated by a wall or diaphragm. Then he imagined an external agent or demon  as ``a finite being who knows the paths and velocities of all the molecules by simple inspection but 
who can do no work except open and close a hole in the diaphragm by means of a slide without mass''. In a letter to his friend Tait, dated December 11th 1867,  Maxwell describes for the first time the operation of the demon:

\begin{quotation}
Let him first observe the molecules in $A$ and when he sees one coming the square of whose velocity is less than the mean square velocity of the molecules in $B$, let him open the hole and let it go into $B$. Next, let him watch for a molecule of $B$, the square of whose velocity is greater than the mean square velocity in $A$, and when it comes to the hole let him draw the slide and let it go into $A$, keeping the slide shut for all other molecules. Then the number of molecules in $A$ and $B$ are the same as at first, but the energy in $A$ is increased and that in $B$ diminished, that is, the hot system has got hotter and the cold colder and yet no work has been done, only the intelligence of a very observant and neat-fingered being has been employed.
\end{quotation}

\begin{figure}
	\centering
	\includegraphics[width=5cm]{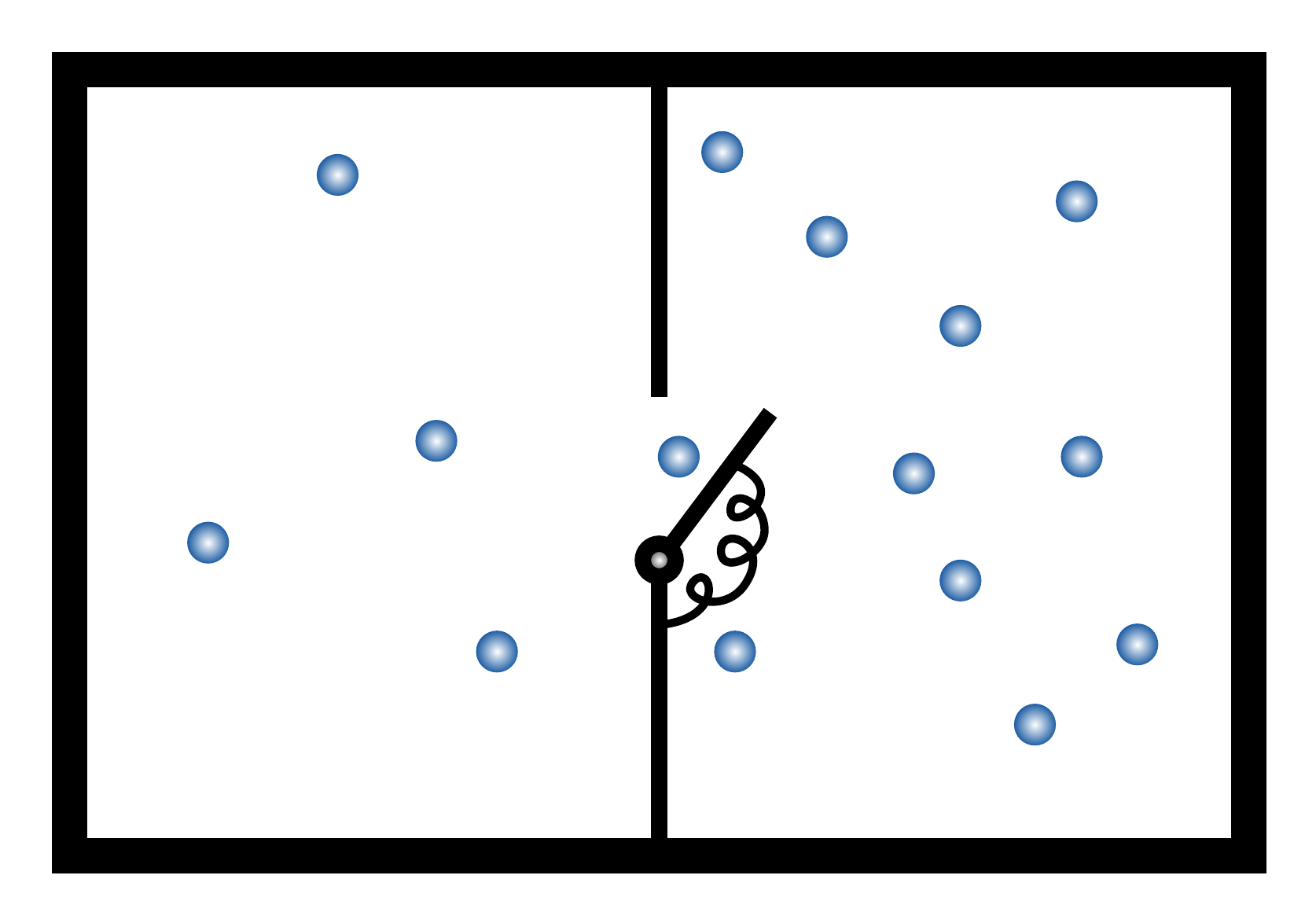}
\caption{The autonomous pressure Maxwell demon.}
\label{fig_pressuredemon2}
\end{figure}



Maxwell 
published the idea, four years after the letter to Tait, in a section of his book, {\em Theory of Heat} (1871). The 
section is entitled ``Limitation of the second law of thermodynamics". Later on, Lord Kelvin coined the name  Maxwell's demon for this ``neat-fingered being'', and it has been subject of 
constant controversy until nowadays \cite{Leff}. The original Maxwell demon opened at least two important lines of research.

The first one is based on a simpler version of the Maxwell demon, also known as {pressure demon}. In this variant, the demon opens the door only when a particle from the left half of the container is moving to the right half, that is, he  allows particles to cross only from left to right. By doing so, the demon 
accumulates particles in the rightmost half of the container, increasing its density  and  inducing a pressure difference that can be subsequently 
used to obtain work. Notice that, in principle,  a simple valve, like the one depicted in Fig.~\ref{fig_pressuredemon2}, could do the same job as the demon. However, this naive idea does not work because the valve is subjected to thermal fluctuations that randomly open the gate and allow particles to cross in the ``wrong'' direction. In fact, there is a fundamental  impossibility of  rectifying thermal fluctuations  due to the reversibility of microscopic dynamics \cite{Ehrich2019}.  The celebrated 
 Smoluchowsky-Feynman ratchet  \cite{Feynman,Smoluchowski:1912} is another example of autonomous Maxwell demon, which cannot rectify thermal fluctuations from a single bath. Feynman went further and assumed that the ratchet  is  immersed in a fluid at a 
temperature lower than the source of the fluctuations and proved that then the 
ratchet is indeed able to rectify and perform work. Nevertheless, the 
second law is not defeated since now the system is in contact with  two thermal baths at different temperatures and the whole setup 
acts as a thermal machine. Feynman showed that the 
efficiency of such a motor cannot be higher than the Carnot efficiency, in agreement with the second law\footnote{Feynman also proved that 
the thermal ratchet can reach Carnot efficiency in the limit of zero power. However, this result has been subject to some criticism:  an autonomous thermal machine can reach Carnot 
efficiency only in the case of tight coupling between heat flow and mechanical work, something that it is 
not possible to reach in the original setup by Smoluchowsky and Feynman \cite{Parrondo1996}.}. This approach can be applied to generic classical systems in contact with thermostats and/or chemostats. This idea has yielded a fruitful line of research on Brownian motors \cite{Reimann:2002hs}, systems that exhibit systematic motion or perform work using gradients of temperature and/or chemical potential, with applications in biology  and nanotechnology \cite{Bustamante2001,Feng2021}. The role of information in Brownian motors is not completely clear, although we introduce in section \ref{sec:flows} the concept of information flows in autonomous systems, which  allows one to analyze some of these machines as an exchange between information and entropy \cite{Allahverdyan2009,Horowitz2014}.

The second line of research directly concerns the use of information by the original Maxwell's demon \cite{Leff,Parrondo2015,Sagawa2012b,sagawa2012thermodynamics}.  Two questions immediately arise form the fact that the demon can beat the second law using information.  First, can we derive a quantitative relationship between the entropy decrease and the information that the demon possesses? Or, in other words, how the acquisition of information modifies the second law? Second, can we restore the validity of the second law by finding an entropy cost associated to the acquisition and/or manipulation of information by the demon? Thermodynamics of information addresses these two questions. However, it is convenient to analyze them separately. We discuss the first question  in section \ref{sec:feedback} and the second one in section \ref{sec:restoring}.

 \begin{figure}
 \begin{center}
 \[
 \includegraphics[width=7cm]{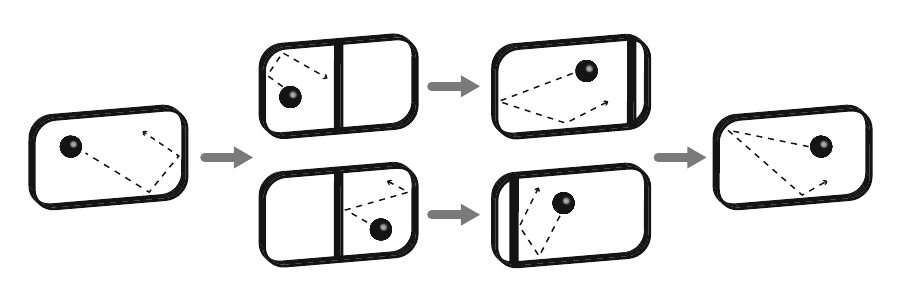}
 \]
 \caption{The Szil\'ard engine.}
 \label{fig_szilard}
 \end{center}
 \end{figure}

In 1929, Szil\'ard introduced a simplified version of the Maxwell demon known as the Szil\'ard engine \cite{Szilard1929}, which is more suitable for the resolution of these issues. It consists of a single-molecule gas  at temperature $T$ in a box of volume $V$. This means that the molecule thermalizes at temperature $T$ in any collision with the wall, i.e., the outgoing velocity is a random variable distributed according to the corresponding equilibrium distribution. An external agent or demon performs the cyclic operations sketched in Fig.~\ref{fig_szilard}. First, he inserts a piston at the middle point of the container and measures in which of the two halves the particles lies. With this information, the demon can realize a reversible isothermal expansion by opposing a force to the pressure $P$ exerted by the single-molecule gas. In the expansion, the demon  extracts a work
\begin{equation}\label{workszilard}
    W_{\rm extract}=\int_{V/2}^VPdV'=\int_{V/2}^V\frac{kT}{V'}dV'=kT\ln 2
\end{equation}
where we have used the equation of ideal gases, $PV'=NkT$,  with $N=1$, and $k$ is the Boltzmann constant.
Finally, the piston is removed and the cycle is completed.  In principle, the insertion and removal of the piston can be performed without any energy expenditure\footnote{This is not true for a quantum particle at low temperature because, due the the Heisenberg uncertainty principle, the kinetic energy increases when it is confined in one of the halves \cite{Zurek1984}.}. Hence, the Szil\'ard engine is able to extract energy from a single thermal bath in a cycle, in apparent contradiction to  the second law. 

Some aspects of the original Szil\'ard cycle are not completely clear, specially the application of the ideal gas equation to the pressure on the piston, which is due to single collisions. Nevertheless, these issues can be resolved and, moreover, the Szil\'ard engine can be implemented in any system exhibiting a symmetry breaking, like an Ising model \cite{Parrondo2001}. In fact, there are  experimental realizations of the Szil\'ard engine in a variety of systems: single-electron circuits \cite{Koski13786,pekola2019}, Brownian particles in optical traps \cite{Roldan2014}, or Brownian rotors in electrostatic potentials \cite{Toyabe2010}.

The Szil\'ard engine, as the original Maxwell demon, utilizes the information gathered in the measurement to decrease the entropy of the system. While in the Maxwell demon the information involved and the decrease of entropy are difficult to assess, in the Szil\'ard engine both have a precise value:  information is acquired in a single  measurement per cycle with two equiprobable outcomes, left ($L$) or right ($R$), and the entropy decreases by  $k\ln 2$. This is why most of the studies on the connection between thermodynamics and information are based on the Szil\'ard engine. In fact, the Szil\'ard engine partially solves the first question mentioned above. Before the measurement, the particle can be in any of the two halves and, after the measurement, it occupies only one. The measurement is equivalent to a compression from a volume $V$ to $V/2$, which  reduces the entropy of the gas by $k\ln 2$. Such a compression would need a work $kT\ln 2$, counterbalancing the work \eqref{workszilard} extracted in the expansion, but  apparently the measurement is able to perform the compression without any energy expenditure.

Thermodynamics of information refines and generalizes this argument \cite{Parrondo2015}. First, by incorporating concepts from information theory that quantify the information acquired in a measurement in a more precise and general way. Second, by realizing that the states resulting from measurements and other processes involving information are in general non-equilibrium states. Hence, thermodynamics of information can be considered as a branch of statistical mechanics that deals with a special class of non-equilibrium states bearing information.

The outline of this article is the following.
In section \ref{sec:info}, we review the concepts of information theory relevant to thermodynamics, like Shannon entropy and mutual information. We discuss in detail
the relationship between Shannon entropy and thermodynamic entropy in section \ref{sec:shantherm}. In section \ref{sec:feedback}, we  show how the acquisition of information affects the second law. The resulting modified second law for feedback processes ---those that follows a protocol depending on the outcome of a measurement--- is the answer to the first question that we have posed above. 
In section \ref{sec:landauer}, we derive the thermodynamic cost of some processes involving information in memories and other information devices, like the celebrated Landauer principle \cite{Landauer1961}. These costs help us to solve
the second question, i.e., the restoration of the original second law in the Szil\'ard engine and any generic isothermal feedback processes. This is discussed in section \ref{sec:restoring}, where we show that the  feedback processes obey the original second law when one accounts for the cost of measurement and/or restoring the demon's memory, generalizing Bennett's analysis of the Szil\'ard engine \cite{Bennett1982b}.
In section \ref{sec:flows} we introduce a formalism that allows one to study information flows \cite{Allahverdyan2008,Horowitz2014} in  autonomous  Maxwell demons. Finally, in section \ref{sec:conclusions} we present our main conclusions and discuss the physical nature of information.

\section{Information}
\label{sec:info}

In his seminal papers published in 1948 \cite{shannon}, 
Shannon introduced a measure of the uncertainty or ignorance that one has about a random object or discrete random variable  $X$, which can be a generic message in a communication channel, a file in a computer, or the micro-state of a physical system. Shannon uncertainty, also known as Shannon entropy\footnote{
In a famous conversation between Shannon and Von Neumann, the latter suggested Shannon to use the name entropy because \eqref{shan} is similar to the definition of entropy in statistical mechanics and, more importantly, because ``nobody really knows what entropy is''. The conversation was reported by Shannon himself but later on he admitted that it probably never took place \cite{oral_historyclaude_2021}.
},
is defined as
\begin{equation}\label{shan}
S(X)=S[p_X]=-\sum_{x} p_{X}(x)\log p_{X}(x).
\end{equation}
Here we have introduced two useful equivalent notations, namely, expressing the dependency of the uncertainty $S$ on the random variable $X$ or on its probability distribution $p_X(x)$. 
Uncertainty can be expressed in bits, if the logarithm in Eq.~\eqref{shan} is in base 2, in nats, if it is the natural logarithm, or in  units  of thermodynamic entropy if we multiply nats by, for instance, the Boltzmann constant $k$. In the rest of the article, we will assume that the Shannon entropy is expressed in physical units of energy divided by temperature, without the need of writing explicitly the Boltzmann constant. Finally, for continuous random variables, the sum in \eqref{shan} must be replaced by an integral \cite{Cover}.

It is customary to interpret Shannon uncertainty as a measure of information, and this is indeed appropriate in certain situations. For example, when we consider the information content of a given instance of the random variable $X$. However, when we talk about information, specially in physics, or, more precisely, about obtaining information from a system, we have in mind an inquiry  about the state $X$ of the system, which consists in  the measurement of a quantity $Y$. When we measure $Y$, we acquire information about $X$ that results in a decrease of its uncertainty. This decrease is precisely the amount of information that $Y$ provides about $X$:
\begin{equation}\label{mutual0}
I(X;Y)\equiv S(X)-S(X|Y)
\end{equation}
which is called {\em mutual information} \cite{Cover,shannon}. Here  $S(X|Y)$ is the uncertainty of the posterior probability distribution $p_{X|Y}(x|y)$ averaged over the possible outcomes $p_{Y}(y)$, i.e.
\begin{eqnarray}
S(X|Y) &\equiv &\sum_{y} p_{Y}(y)\left[-\sum_{x} p_{X|Y}(x|y)\log p_{X|Y}(x|y)\right]\nonumber
\\
&=&-\sum_{x,y} p_{XY}(x,y)\log p_{X|Y}(x|y).
\end{eqnarray}
Using the relationship between conditional and joint probabilities (Bayes' formula): $p_{XY}(x,y)=p_{X|Y}(x|y)p_{Y}(y)$, we can write the mutual information as
\begin{equation}\label{mutuali0}
I(X;Y)=\sum_{x,y} p_{XY}(x,y)\log \frac{p_{XY}(x,y)}{p_{X}(x)p_{Y}(y)}.
\end{equation}

From this expression we can derive a number of interesting properties of the mutual information. First,  mutual information is symmetric, i.e., $Y$ provides the same information about $X$ as  $X$ provides about $Y$. Second, using the properties of the logarithm, one can prove from Eq.~\eqref{mutuali0} that $I(X;Y)$ is always positive and vanishes if and only if $X$ and $Y$ are statistically independent \cite{Cover}. Hence,  mutual information is a measure of the correlation between $X$ and $Y$. Third, if we measure without errors a quantity $Y=f(X)$, then $S(Y|X)=0$ and $I(X;Y)=S(Y)$. In other words, the information provided by an error-free measurement is the uncertainty of the outcome.

Finally, Eq.~\eqref{mutuali0} allows us to write the mutual information in three different ways:
\begin{eqnarray}
I(X;Y) &=& S(X)-S(X|Y) \nonumber \\
&=& S(Y)-S(Y|X) \nonumber \\
&=& S(X)+S(Y)-S(X,Y)\geq 0 \label{mituali1}
\end{eqnarray}
where $S(X,Y)$ is the Shannon entropy or uncertainty of the joint probability distribution $p_{XY}(x,y)$.
The last equality indicates that correlations make the entropy sub-additive:
\begin{equation}
\label{mituali2}
S(X,Y)=S(X)+S(Y)-I(X;Y).
\end{equation}
This expression will be useful when we consider the physical nature of Maxwell or Szil\'ard demons and interpret a measurement as the creation of correlations between the state of the demon and the state of the system.

\section{Shannon and thermodynamic entropies}
\label{sec:shantherm}

If Shannon uncertainty is identified  as thermodynamic entropy, one immediately obtains from the definition of mutual information \eqref{mutual0} that the entropy of a system $X$ decreases by an amount $I(X;M)$ when we measure a quantity  $M$. For instance, in the case of the Szil\'ard engine with error-free measurements,  $X$ is the micro-state of the global system, particle plus bath, which obviously includes the position of the particle, and $M=L,R$ ($L$ for left, and $R$ for right) is the outcome of the measurement, i.e., the half of the container where the particle lies after the insertion of the piston. Since $M$ is a function of $X$ and the two outcomes are equally probable, $I(X;M)=S(M)=k\ln 2$. Thus, the total entropy  decreases by $k\ln 2$, and this decrease allows one to extract an energy $kT\ln 2$ from the thermal bath in the form of work by driving quasi-statically the system back to its initial state. This is a similar argument as the one that we sketched in section \ref{sec:intro}. 

This explanation, however, requires a  more careful consideration. The identification of Shannon and thermodynamic entropies is not correct, in general.
The main reason is that the Shannon entropy $S[\rho(x,t)]$ of the probabilistic state  of a  classical system is invariant under Hamiltonian evolution. Here $x$ denotes a micro-state of the system, i.e., the value of the positions and momenta of all its particles, and $\rho(x,t)$ is the probability density of observing a micro-state $x$ at time $t$. The same argument holds for quantum systems, replacing $\rho(x,t)$ by the density matrix $\rho(t)$ and the Shannon entropy by Von Neumann entropy, ${\rm Tr}(\rho\ln\rho)$ \cite{schumacher2010}. Here, we will restrict our discussion to classical systems, although it is not difficult to extend it to quantum systems \cite{Esposito2009}.

To observe an increase of Shannon entropy, as dictated by the second law  for irreversible processes, one has to consider information losses that degrade the probabilistic state $\rho(x,t)$, like  coarse-graining or the inaccessibility of detailed information about certain degrees of freedom. 

A second important property of thermodynamic entropy is its connection with the energetics of a process through  Clausius equation, which relates the temperature $T$ to the increase of the entropy, $\delta S$, and the energy, $\delta E$, of a system\footnote{Clausius originaly introduced this equation as a definition of entropy. However, it is in fact a definition of temperature for systems at equilibrium, since energy and entropy are more fundamental magnitudes that can be obtained from  dynamical properties, like the Hamiltonian and the phase space volume of regions of constant energy.}: 
\begin{equation}\label{clausius}
\delta S=\frac{\delta E}{T}.
\end{equation}
This equation is crucial to find the limitations that the second law imposes on  energy extraction from thermal reservoirs.  For instance, if a  system is in contact with a thermal bath at temperature $T$, according to Eq.~\eqref{clausius} the total entropy production in a process is 
\begin{equation}\label{2ndlaw}
\Delta S_{\rm tot}=\Delta S-\frac{Q}{T}\geq 0
\end{equation}
where $\Delta S$ is the change of entropy in the system and $Q$ is the energy transferred from the bath to the system. If an external agent manipulates the system preforming a work $W$, the change of energy in the system is $\Delta E=Q+W$ and
\begin{equation}
T\Delta S_{\rm tot}=T\Delta S-\Delta E+W=W-\Delta F
\end{equation}
$\Delta F$ being the increase of free energy, $F=E-TS$, in the system. The second law $\Delta S_{\rm tot}\geq 0$ for this isothermal process is equivalent to
\begin{equation}\label{2ndlawfw}
W\geq \Delta F.
\end{equation}
In particular, for a cycle $\Delta F=0$ and the inequality, $W\geq 0$,  tells us that it is impossible to extract energy from a thermal bath in a cyclic process.

For a system in contact with equilibrium thermal reservoirs, one can solve these two issues ---namely, the invariance of Shannon entropy under Hamiltonian evolution and its connection with the energetics of a process--- with the two following assumptions that specify the  information available to an external observer 
\cite{Esposito2009}: first, the observer does not have access to the correlations between the system and the baths and, second, the only information about the state of the baths is their average energy. More precisely, the baths are described by thermal states with  given average energies. In the case of a single bath with micro-states $z$, if the actual probabilistic state of the global system at time $t$ is $\rho(x,z;t)$, this loss of information yields an effective state $\rho_{\rm obs}(x,z;t)$ given by
\begin{equation}\label{degrad}
\rho(x,z;t)\to \rho_{\rm obs}(x,z;t)=\rho(x,t)\,\rho_{\rm B, eq}(z;T(t))
\end{equation}
where $\rho(x,t)$ is the marginal probabilistic state of the system
\begin{equation}
\rho(x,t)=\int dz\,\rho(x,z;t)    
\end{equation}
and $\rho_{\rm B, eq}(z;T)$ is the thermal state of the bath
\begin{equation}\label{gibbs}
\rho_{\rm B, eq}(z;T)=\frac{e^{-\beta H_{\rm B}(z)}}{Z_{\rm B}(\beta)}.
\end{equation}
Here $H_{\rm B}(z)$ is the Hamiltonian of the bath, $\beta=1/(kT)$  the inverse temperature, and $Z_{\rm B}(\beta)$  the partition function. The temperature $T(t)$ in \eqref{degrad} is set as the one that verifies 
\begin{eqnarray}
E_{\rm B}(t) &\equiv &
\int dxdz\,H_{\rm B}(z)\,\rho(x,z;t)
\nonumber \\
&=&\
\int dz\,H_{\rm B}(z)\,\rho_{\rm B, eq}(z;T(t)),
\label{ener}
\end{eqnarray}
i.e., both distributions have the same average bath energy $E_{\rm B}(t)$  at any time $t$. 

The replacement of the actual state $\rho(x,z;t)$ by the observable state $\rho_{\rm obs}(x,z;t)$ solves the two aforementioned issues: the invariance of Shannon entropy under Hamiltonian evolution and the relationship between entropy and energy given by the Clausius equation \eqref{clausius}. First, applying \eqref{mituali2}, we find that
the replacement \eqref{degrad} increases the Shannon entropy:
\begin{equation}\label{sobs}
S[\rho_{\rm obs}(x,z;t)]-S[\rho(x,z;t)] =I(X;Z)+\Delta S_{\rm B} \geq 0
\end{equation} 
where $\Delta S_{\rm B}=S[\rho_{\rm B, eq}(z;T(t))]-S[\rho(z;t)]$. The inequality follows from the positiveness of the mutual information and the fact that the thermal sate \eqref{gibbs} is the one that maximizes the Shannon entropy under condition \eqref{ener}. 

Consider  a process where we manipulate the system, whose Hamiltonian $H(x;t)$ is now time-dependent. We also assume that initially system and bath are uncorrelated and the latter is at equilibrium, i.e., $\rho(x,z;0)=\rho(x,0)\rho_{\rm B,eq}(z;T(0))$. Then, $\rho(x,z;0)=\rho_{\rm obs}(x,z;0)$ and, since  Shannon entropy is invariant under the Hamiltonian evolution, $S[\rho(x,z;t)]=S[\rho_{\rm obs}(x,z;0)]$. Hence, from Eq.~\eqref{sobs}, we obtain the entropy production in the process:
\begin{equation}\label{sprod1}
S[\rho_{\rm obs}(x,z;t)]-S[\rho_{\rm obs}(x,z;0)] =I(X;Z)+\Delta S_{\rm B} \geq 0.
\end{equation} 
This expression reflects the two information losses that we have introduced in \eqref{degrad}: $I(X;Z)$ is the information associated to the correlations between the system and the bath, whereas $\Delta S_{\rm B}$ is the increase of Shannon entropy that results from replacing the actual state of the bath by a thermal state \cite{Esposito2009}. Each of these terms is positive and we recover a second law for the Shannon entropy $S[\rho_{\rm obs}(x,z;t)]$. It is interesting to notice that both terms can be relevant in realistic situations, as shown in \cite{Ptaszynski2019}.

Alternatively, the entropy production can also be calculated by using the explicit form of the thermal state \eqref{gibbs}:
\begin{equation}
    S[\rho_{\rm obs}(x,z,t)] =S[\rho(x,t)]+k\beta(t) E_{\rm B}(t) +k\ln Z_{\rm B}(\beta(t)).
\end{equation}
Since $E_{\rm B}=-\partial \ln Z_{\rm B}/\partial\beta$, the time derivative of the Shannon entropy of the observable state can be written as
\begin{equation}\label{clausiusprocess}
\dot S[\rho_{\rm obs}(x,z,t)] = \dot S[\rho(x,t)]+\frac{\dot E_{\rm B}(t) }{T(t)}.
\end{equation}
Integrating over the whole process with the initial condition discussed above, and taking into account that the bath is large enough so that the change of temperature is negligible\footnote{Notice that we have to consider a time-dependent temperature to derive the variation of Shannon entropy in the bath in Eq.~\eqref{clausiusprocess}. However, this variation can be neglected when we integrate this equation, if the heat capacity of the bath is large enough.}, we finally get
\begin{equation}\label{sprod2}
S[\rho_{\rm obs}(x,z;t)]-S[\rho_{\rm obs}(x,z;0)] =\Delta S-\frac{Q}{T} 
\end{equation} 
where $Q=E_{\rm B}(0)-E_{\rm B}(t)$ is the net energy transferred from the bath to the system. From Eq.~\eqref{sprod1}, we conclude that the entropy production, as expressed in Eq.~\eqref{sprod2}, is positive.
Thus, we recover the same expression for the entropy production as in \eqref{2ndlaw}, i.e., 
the loss of information implied by \eqref{degrad} is compatible with Clausius equation \eqref{clausius}.  The standard thermodynamic argument to derive \eqref{2ndlawfw} can now be applied to the {\em non-equilibrium free energy}
\begin{equation}\label{noneqfe}
    {\cal F}(\rho,H)\equiv\langle H(x)\rangle-TS[\rho(x)]
\end{equation}
yielding a bound to the work needed to complete an isothermal process:
\begin{equation}\label{secondlawnoneq}
   W \geq  \Delta {\cal F}.
\end{equation}
This can be considered an extension of the second law \eqref{2ndlawfw} for isothermal processes involving non-equilibrium states, and  reduces to the standard second law if the initial and final states are equilibrium, since the non-equilibrium free energy \eqref{noneqfe} is equal to the standard equilibrium one for thermal states like \eqref{gibbs}. 
Alternatively to our derivation based on information losses, \eqref{secondlawnoneq} can be directly obtained from a master equation obeying local detailed balance \cite{Esposito2011}, which is the equation that describes  systems in contact with reservoirs in the Markovian approximation.

 The bound  \eqref{secondlawnoneq} is saturated for operationally reversible process, i.e., for processes where the system visits the same states as in the forward process when the protocol is reversed  \cite{Parrondo2015}.
 This observation puts into question the utility of \eqref{secondlawnoneq} for generic  non-equilibrium states, since any  irreversible relaxation to equilibrium prevents  the bound to be reached. One strategy to overcome this issue consists in modifying instantaneously the Hamiltonian to convert the initial and final states into equilibrium states, although this protocol is a bit artificial and sometimes unrealizable \footnote{Any state $\rho(x)$ can be converted into an equilibrium state by  setting the Hamiltonian equal  to $H(x)=-kT\ln\rho(x)$. Notice however that such Hamiltonian can be difficult or impossible to realize as, for instance, when $\ln\rho(x)$ is not a quadratic function of velocities.}  \cite{Parrondo2015}. However, the bound \eqref{secondlawnoneq} can be tight as well for systems with slow and fast degrees of freedom. If there is a large separation of time scales and the fast degrees of freedom are equilibrated during the process, then the inequality \eqref{secondlawnoneq} 
is met. As we will see in section \ref{sec:landauer}, it turns out that this is the case of many relevant processes involving information.

\section{Second law for feedback processes}
\label{sec:feedback}

In an isothermal feedback process, like the Szil\'ard engine, the external agent measures a magnitude $M$ at a given time $t_{{\rm m}}$ and uses this information to complete an isothermal process. The measurement is characterized by the conditional probability  $p_{M|X}(m|x)$, where $m$ is the measurement outcome and $x$ is the microscopic state of the system. This includes error-free measurements,  where  $p_{M|X}(m|x)=1$ if $m=m(x)$ and zero otherwise, but it can also describe measurements affected by noise or any other source of inaccuracy. Using the measurement outcome, the observer  updates the probabilistic state of the system $\rho_X(x,t_{\rm m})$ according to Bayes rule
\begin{equation}\label{bayesupdate}
\rho_X(x,t_{\rm m})\to   \rho_{X|M}(x|m)=\frac{p_{M|X}(m|x)\rho_X(x,t_{\rm m})}{p_M(m)}.
\end{equation}
Here we are assuming the most common case, where $x$ is a continuous random variable and $m$ is discrete, and  use the Latin letter $p$ for probability distributions and the Greek one $\rho$ for probability densities. Nevertheless, the argument can be easily generalized to other cases. For a given outcome $m$, the updated state $\rho_{X|M}(x|m)$ is in general a non-equilibrium state, even if the  state before the measurement, $\rho_{X}(x,t_{\rm m})$, is equilibrium.
We see here the first appearance of non-equilibrium states in processes involving information. The unconditional state of the system after the measurement is again
\begin{equation}\label{rhoxafter}
    \sum_m p_M(m)\rho_{X|M}(x|m)=\rho_X(x,t_{\rm m}),
\end{equation}
i.e., the measurement does not alter the system. In section \ref{sec:restoring}, we will establish this as a condition for an ideal classical measurement. Given an outcome $m$, the free energy after the measurement is
\begin{align}
    &{\cal F}[\rho_{X|M}(x|m),H(x,t_{\rm m})]\nonumber \\&=\int dx H(x,t_{\rm m})\rho_{X|M}(x|m)-TS[\rho_{X|M}(x|m)]
\end{align}
where $H(x,t)$ is the Hamiltonian of the system at time $t$.
Averaging over the possible outcomes we get the non-equilibrium free energy after the measurement
\begin{align}
    {\cal F}_{\rm post}&= \sum_m p_M(m){\cal F}[\rho_{X|M}(x|m),H(x,t_{\rm m})]\nonumber \\
    &=\int dx H(x,t_{\rm m})\rho(x,t_{\rm m})-TS(X|M).
\end{align}
Using the definition of mutual information \eqref{mituali2}, we obtain that, as a consequence of the Bayes update, the non-equilibrium free energy of the system changes as
\begin{eqnarray}\label{mes0}
    \Delta {\cal F}_{\rm meas} &\equiv & {\cal F}_{\rm post} -{\cal F}_{\rm pre}\nonumber\\
    &=& T[(S(X)-S(X|M)]\nonumber\\
    &=& TI(X;M),
\end{eqnarray}
where ${\cal F}_{\rm pre}={\cal F}[\rho(x,t_{\rm m}),H(x,t_{\rm m})]$ is the non-equilibrium free energy of the system immediately before the measurment.

\begin{figure}
\begin{center}
\includegraphics[width=5cm]{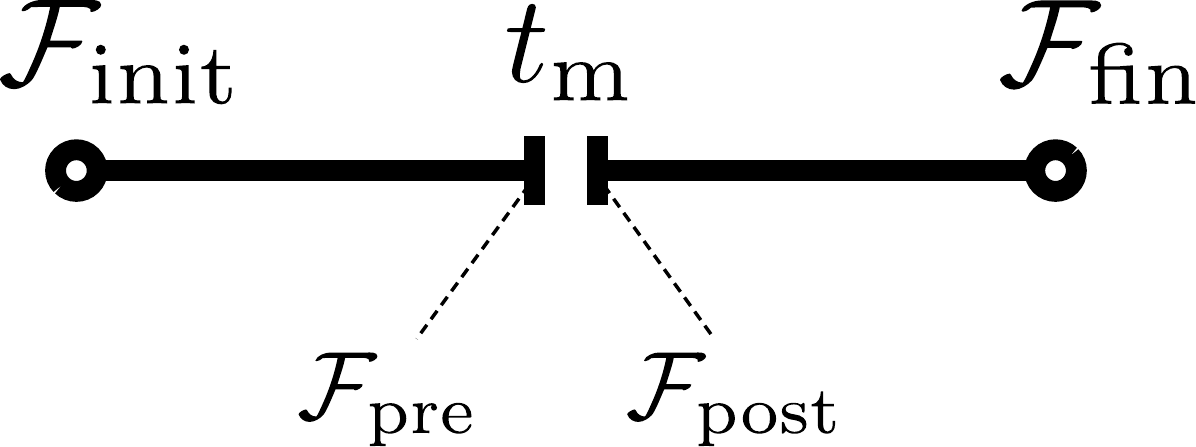}
\caption{Scheme of a basic feedback process with a single measurement.}
\label{fig:process_fb}
\end{center}
\end{figure}

We can now apply Eq.~\eqref{secondlawnoneq} to the subprocesses that take place before and after the measurement. The total work in the feedback process is bound as
\begin{eqnarray}
W_{\rm fb}&\geq & {\cal F}_{\rm fin}-{\cal F}_{\rm post}+{\cal F}_{\rm pre}-
{\cal F}_{\rm init}\nonumber \\
&=&
\Delta {\cal F} -\Delta{\cal F}_{\rm meas}\label{mes1}
\end{eqnarray}
where $\Delta {\cal F}={\cal F}_{\rm fin}-{\cal F}_{\rm init}\,$ and the subscripts, init, pre, post, fin, indicate respectively the states at the different stages of the process: initial and final, and immediately before and after the measurement (see Fig.~\ref{fig:process_fb}).
Combining \eqref{mes0} and \eqref{mes1}, we finally obtain
\begin{equation}\label{2ndlwafb}
W_{\rm fb}\geq \Delta {\cal F} -TI(X;M)
\end{equation}
which is the second law for feedback processes. This result was derived by Sagawa and Ueda for classical and quantum systems \cite{Sagawa2008}, although the relevance of mutual information in thermodynamics was first pointed out by Lloyd and Touchette \cite{Lloyd1989,Touchette2000}. It indicates that the work necessary to complete a process decreases by an amount $TI(X;M)$ if we use the information gathered in the measurement. In particular, for a cycle the work can be negative, i.e., we can extract a work $W_{\rm extract}=-W_{\rm fb}$ at least equal to $TI(X;M)$. The Szil\'ard engine, where $I(X;M)=k\ln 2$ is an explicit example that saturates the bound.  

Notice that, due to \eqref{rhoxafter}, the unconditional  non-equi\-li\-brium free energy of the system does not change in the measurement. The mutual information $I(X;M)$ in \eqref{mes0} and \eqref{2ndlwafb} appears because we average the conditional entropy over $p_M(m)$. The reason of averaging in this way is that, in a feedback process, the protocol depends on the outcome $m$, and so does the work. The bound \eqref{2ndlwafb} applies to the average work $W_{\rm fb}$ over all possible outcomes and protocols, each one obeying \eqref{secondlawnoneq}. If the information provided by the measurement is not used in the protocol from $t_{\rm m}$ to the final time of the process, then we recover the standard second law \eqref{secondlawnoneq} with equilibrium free energies.

The bound \eqref{2ndlwafb} does not resolve the problem of reconciling the Szil\'ard engine with the original second law. However, it is an important result because it establishes a benchmark to the work that can be extracted in a feedback process. It also indicates that  the mutual information $I(X;M)$ is a thermodynamic resource and can be used to define
 efficiencies for information and hybrid engines \cite{Saha2021b,Schmitt2015} and to compare the performance of information motors and of chemical or thermal machines \cite{Horowitz2013}. It is also interesting to explore which protocols saturate the bound \eqref{2ndlwafb}. It is found that this happens for operationally reversible processes, those which visit the same states when the action of the external agent is reversed \cite{Horowitz2011b}. Notice however that the notion of reversibility of feedback processes requires a more careful discussion since it is not trivial to define the time reversal of a measurement \cite{Horowitz2011,Horowitz2010}.

The second law \eqref{2ndlwafb} can be further refined by deriving fluctuation theorems. These theorems are exact equalities that hold for any process even far from equilibrium, and from which the second law can be obtained as a corollary \cite{Horowitz2010,Ponmurugan2010,Sagawa2009,Sagawa2010,Sagawa2012}.

\section{Informational states and Landauer's principle}
\label{sec:landauer}

Information devices, like memories, processors, or the biochemical machinery of DNA and RNA, are systems that can adopt different informational states that have a long lifetime and are interchangeable. These states are, for example, the bits in a hard drive, which can be 0 or 1, or the bases of DNA, which can be any of the four nucleotides:  cytosine, guanine, adenine or thymine.

\begin{figure*}
\begin{center}
\includegraphics[width=14cm]{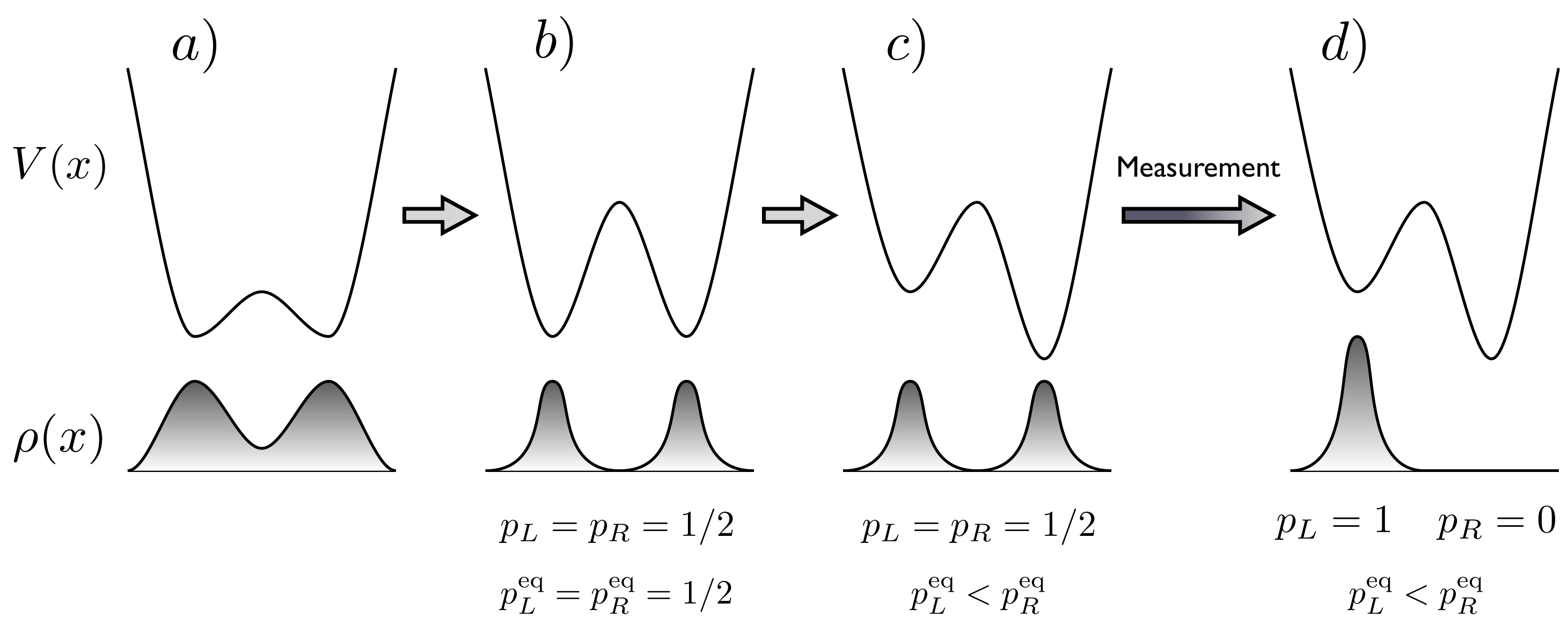}
\caption{Informational states of a Brownian degree of freedom in a bistable potential. The  rise of the barrier from  a) to b) illustrates the creation of information \cite{Roldan2014}, whereas c) and d) show non-equilibrium probabilistic informational states, which are the result of  the history and of a measurement, respectively.}
\label{fig:info_states}
\end{center}
\end{figure*}

These two properties, long lifetimes and interchangeability, imply some kind of effective symmetry breaking in the system: its phase space is partitioned into several regions where the system remains for a very long time. Each region is an informational state and the system is no longer ergodic in the whole phase space. Here, we refer to an effective lack of ergodicity when there is a huge separation between the time scales of the dynamics within the regions and of the jumps between regions. This type of effective ergodicity breaking occurs due to phase transitions in finite systems or to the presence of high energy barriers with a negligible probability to be crossed.
A paradigmatic example is a Brownian degree of freedom in a double well potential, like the one depicted in figure \ref{fig:info_states}.

In general, the phase space $\Gamma$ of an information device is partitioned into a number of regions $\Gamma_m$, with $m=1,2,\dots,{\cal M}$.
We will assume that the dynamics within each region is very fast and that the system is in local equilibrium within every $\Gamma_m$. On the other hand, the system can be in any of the informational states with a certain probability $p_m$ that depends on the history, measurements, etc. We show an explicit example in figure \ref{fig:info_states}, based on a Brownian particle or degree of freedom in a double well potential. By raising the barrier far above $kT$, we create a memory from a) to b). If the potential is symmetric, each informational state $m=L,R$ is populated with the same probability $p_L=p_R=1/2$ and the system is in global equilibrium. However, if we change the energy of the wells, as in c), populations do not change due to the high barrier but the equilibrium probabilities $p^{\rm eq}_{R,L}$ depart from $1/2$. Hence, the system is no longer in global equilibrium. The informational state can also change due to a measurement, as in the transition from c) to d).

If the device is at temperature $T$, the state of the system is then given by
\begin{equation}\label{rhoinfo}
\rho(x)=\sum_m p_m \frac{e^{-\beta H(x)}}{Z_m}\chi_m(x)
\end{equation}
where $x\in \Gamma$ is a micro-state, $H(x)$ is the Hamiltonian, $\chi_m(x)$ is the index function of  $\Gamma_m$, i.e., $\chi_m(x)=1$ if $x\in\Gamma_m$ and zero otherwise, and $Z_m$ is the restricted partition function
\begin{equation}
{Z_m}\equiv \int_{\Gamma_m}dx\, e^{-\beta H(x)}.
\end{equation}
The probability distribution $\{p_m\}$, with $m=1,2,\dots,{\cal M}$  is arbitrary and defines a probabilistic informational state. The main characteristic of the information device is that it can adopt any probabilistic informational state and that an external agent can drive the system, as in figure \ref{fig:info_states}, from one probabilistic informational state to another,
writing or processing information in the device.
A special probabilistic informational state is the one corresponding to global equilibrium $p_m=Z_m/Z$, where $Z=\sum_mZ_m$ is the global partition function. 

The energetics of processes involving informational states can be analyzed using the non-equilibrium free energy \eqref{noneqfe}, which for  state  \eqref{rhoinfo} reads
\begin{equation}
    {\cal F}(\rho,H)=\sum_m p_m F_m-TS(p_m)
\end{equation}
where $F_m=-kT\ln Z_m$ is the restricted free energy and $S(p_m)$ is the Shannon entropy of the probabilistic informational state. As already mentioned, the non-equilibrium free energy equals the equilibrium one, ${\cal F}(\rho,H)=-kT\ln Z$ if the system is in global equilibrium. For a symmetric memory where $F_m=F_{\rm local}$ is the same for all informational states $m$, we have
\begin{equation}
    {\cal F}(\rho,H)= F_{\rm local}-TS(p_m).
\end{equation}

Now suppose that we manipulate a symmetric memory driving the system isothermally  from a distribution $p_m$ to $p'_m$. If the initial and final restricted free energies are equal, then the work needed to complete the process is bound by 
\begin{equation}\label{2ndlawmanip}
    W\geq T[S(p_m)-S(p'_m)].
\end{equation}
This equation tells us that we have to perform work to order a memory, $S(p'_m)<S(p_m)$, whereas we can extract work by disordering a symmetric memory $S(p'_m)>S(p_m)$.

Landauer's principle is a special case of Eq.~\eqref{2ndlawmanip}. Landauer considered a simple process where a binary symmetric memory is initially in a random state $p_0=p_1=1/2$, with $S(p_m)=k\ln 2$ and is driven to the informational state 0 with probability one, i.e., $p'_0=1$ and $p'_1=0$, with zero Shannon entropy. This process is called {\tt restore-to-zero}, and is usually considered as the erasure of the initial random bit, although it is more appropriate to call it {overwriting}, since the initial unknown bit is replaced by a given one, 0 in our example. The work needed to complete Landauer's overwriting is
\begin{equation}\label{landauer}
    W\geq kT\ln 2
\end{equation}
which is the so-called Landauer's principle. The Landauer principle has been confirmed experimentally in different systems, like Brownian particles in optical traps \cite{Berut2011} (see \cite{Proesmans2020} for a discussion on experimental results on the energetics of overwriting at finite speed). Landauer's overwriting is a special case of a logically irreversible operation, since the initial informational  state of the memory (input) cannot be recovered from the final one (output). It is a common misunderstanding  to interpret Landauer's principle as if logical irreversibility implied thermodynamic irreversibility. This is not true. If the inequality \eqref{landauer} is met, the overwriting process is still logically irreversible but  thermodynamically reversible because the decrease of entropy in the system is compensated by the increase of entropy in the thermal bath due to heat dissipation. Hence, the total entropy production is zero. See \cite{Sagawa2014} for a detailed discussion on the relation between thermodynamic and logical reversibility.

On the other hand, if a memory has a low Shannon entropy, one can extract work by disordering it. The thermodynamics of these information reservoirs that can act as  thermodynamic resources has been analyzed in detail by Barato and Seifert
\cite{Barato2014b}, and by Wolpert  \cite{wolpert2019}. Mandal and Jarzynski have devised   a concrete realization of  work extraction from an ordered memory using  a kinetic model that reads the content of a tape \cite{Mandal2012}.

\section{Restoring the second law}
\label{sec:restoring}

We can now address the question of restoring the original second law  in the Szil\'ard engine by considering the physical nature of the demon, who must possess a memory to  register the measurement outcomes and is subjected to the thermodynamic limitations discussed in the previous section.

First, let us consider the energetics of a measurement. An ideal classical measurement consists of the interaction between a system and a measurement apparatus, which fulfills the following requirements: {\em i)} system and apparatus are initially uncorrelated; {\em ii)} the system is not affected by the interaction; and {\em iii)} system and apparatus do not interact before and after the measurement, i.e., the Hamiltonian of the global system is $H_{\rm sys}(x)+H_{\rm app}(y)$, where $x$ and $y$ are micro-states of the system and the apparatus, respectively. Let $Y$ and $Y'$ be the random variables denoting the state of the apparatus before and after the measurement, respectively, and $p_{M|X}(m|x)$ the
conditional probability  characterizing the measurement. We will further assume that the outcome  is a function of the apparatus micro-state $M=m(Y')$ and that all the information about $X$ is provided by the measurement outcome, i.e., $\rho_{X|Y}(x|y)=\rho_{X|M}(x|m(y))$.

\begin{figure}\label{fig:feedback}
\begin{center}
\includegraphics[width=4cm]{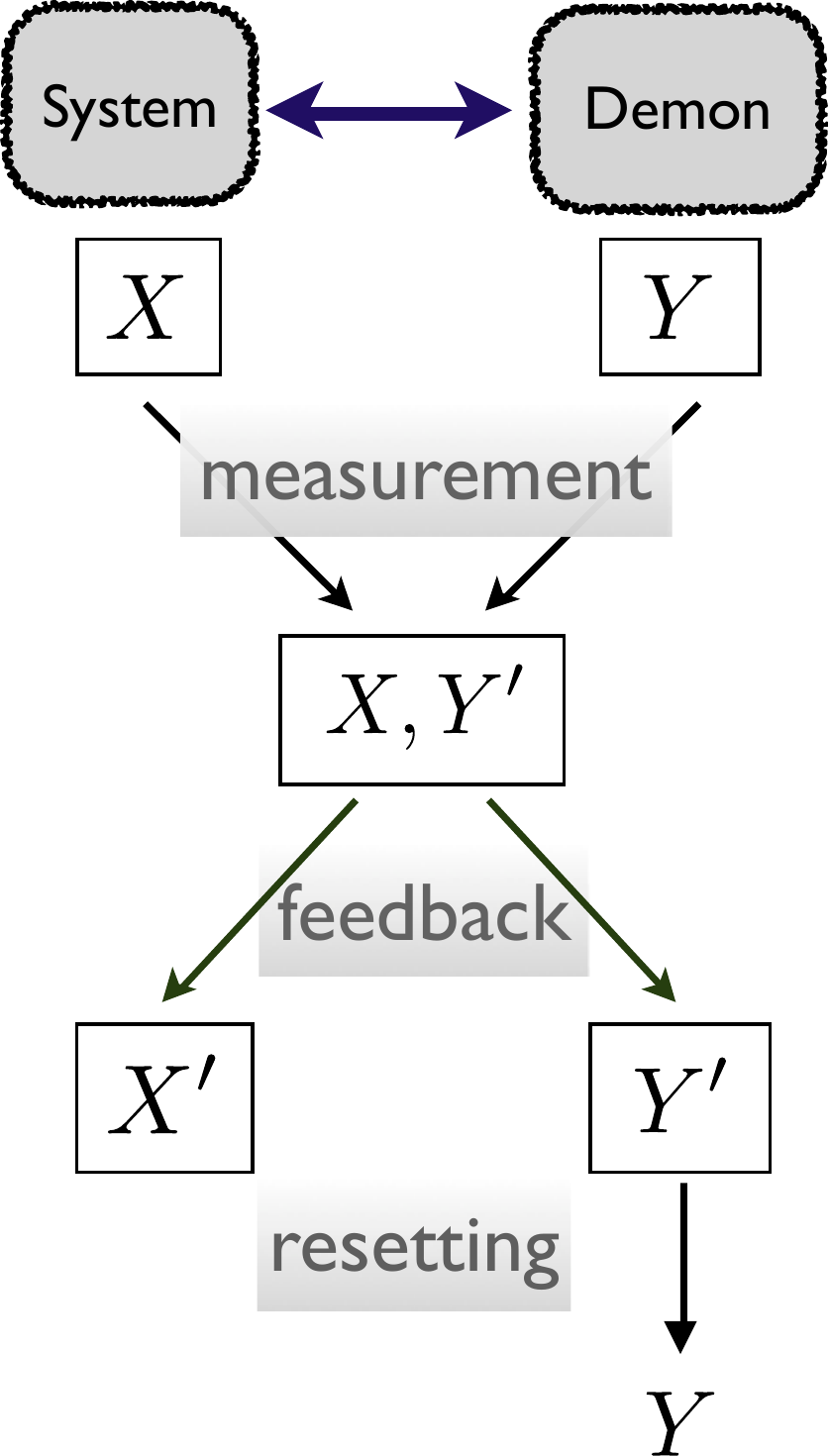}
\caption{Sketch of the operations in a feedback cycle.}
\label{fig:measurement_scheme}
\end{center}
\end{figure}

If the initial state
of the global system immediately before the measurement is
\begin{equation}
\rho_{XY}(x,y)=\rho_X(x)\rho_{Y}(y)
\end{equation}
with non-equilibrium free energy
\begin{equation}
    {\cal F}(X,Y)={\cal F}(X)+{\cal F}(Y),
\end{equation}
then the state after the measurement is
\begin{equation}
\rho_{XY'}(x,y)  = \sum_m \rho_X(x)p_{M|X}(m|x)\rho_{Y'|M}(y|m)
\end{equation}
where $\rho_{Y'|M}(y|m)=\rho_{Y'}(y)/p_M(m)$ if $m = m(y)$ and zero otherwise.
The  marginal probability density for the state of the system  $\rho_X(x)$ does not change a s a consequence of the measurement. This is why we can still denote the random variable corresponding to the state of the system as $X$, whereas the apparatus changes from $Y$ to $Y'$ (see Fig.~\ref{fig:measurement_scheme}).
With these assumptions, it is not hard to prove that $I(X;Y')=I(X;M)$, since  $Y'$ provides  information about $X$ through the outcome $M$. Since the energy of the system does not change due to the measurement, the non-equilibrium free energy of the global system $(X,Y')$ immediately after the measurement can be written as
\begin{eqnarray}
    {\cal F}(X,Y')&=&{\cal F}(X)+{\cal F}(Y')+TI(X;Y') \nonumber\\
    &=&{\cal F}(X)+{\cal F}(Y')+TI(X;M)
\end{eqnarray}
where we have used Eq.~\eqref{mituali2} to express the total Shannon entropy $S(X,Y')$ in terms of the mutual information $I(X;Y')$. 
Hence, the work needed to complete the measurement is
\begin{equation}
    W_{\rm meas}\geq \Delta {\cal F}=\Delta {\cal F}_Y+TI(X;M)
\end{equation}
where $\Delta {\cal F}_Y={\cal F}(Y')-{\cal F}(Y)$.
 Since $I(X;M)$ is positive, we see that measuring or, more generally, creating correlations between two systems, increases the free energy and, if not counterbalanced by $\Delta{\cal F}_Y$, needs work and heat dissipation to be completed. 

As discussed in section \ref{sec:feedback}, the demon can extract  a work $W_{\rm extract}=-W_{\rm fb}$ with $W_{\rm fb}\geq TI(X;M)$ if he uses the information acquired in the measurement in a cyclic process, where the system is driven back to  its initial state $X$.

However, to complete the cycle, the apparatus must be also driven to its initial state $Y$. In doing so, the demon must perform a work
\begin{equation}
    W_{\rm reset}\geq {\cal F}(Y)-{\cal F}(Y')=-\Delta {\cal F}_Y.
\end{equation}
The total work in the process is then
\begin{equation}
    W_{\rm tot}= W_{\rm meas}+ W_{\rm fb}+ W_{\rm reset}\geq 0.
\end{equation}
Hence, the validity of the second law for feedback processes is restored when the entropy costs of measurement and/or resetting the demon's memory are taken into account. This is the generalization of Bennett's analysis of the Szilard engine. Bennett discussed in \cite{Bennett1982b} a realization of the  Szilard engine  where $ W_{\rm meas}=0$ and the demon must overwrite the outcome of the measurement performing a work $kT\ln 2$, as dictated by Landauer's principle. This is achieved if $\Delta {\cal F}_Y=-TI(X;M)$.

From the previous discussion, we see that the Szil\'ard engine can be interpreted as a relatively simple exchange between work and the free energy $TI(X;M)$ stored in the correlations between the system and the demon. For instance, if $\Delta {\cal F}_Y=0$, the engine consists of creating correlations in the measurement, an operation that requires a work $TI(X;M)$ and increases the free energy by the same amount, and then destroying these correlations in the feedback, where the same amount of work is now extracted.

The previous discussion can be presented in terms of entropy, instead of free energy, and extended to continuous time.  Consider now that both the system and the demon are random processes, $X(t)$ and $Y(t)$, respectively.
The time derivative of Eq.~\eqref{mituali2} reads
\begin{equation}
\label{mituali2dt}
\dot S(X(t),Y(t))=\dot S(X(t))+\dot S(Y(t))-\dot I(X(t);Y(t))
\end{equation}
and the total entropy production can be written as
\begin{equation}
\label{totalproddt}
\dot S_{\rm prod}=\dot S(X(t))+\dot S(Y(t))-\dot I(X(t);Y(t))+\dot S_{\rm env}(t)\geq 0
\end{equation}
where $\dot S_{\rm env}(t)$ is the change per unit of time  of the entropy of the environment at time $t$. For instance, if the system and/or the demon are in contact with several thermal baths at different  temperatures $T_i$, then $\dot S_{\rm env}(t)=-\sum_i \dot Q_i(t)/T_i$, where $Q_i(t)$ is the energy transfer per unit of time from the $i$-th bath to the system.

The scheme of figure \ref{fig:measurement_scheme} can be seen as the different stages of a continuous evolution. In the measurement, the system does not change, hence $\dot S(X(t))=0$, and the integration of \eqref{totalproddt} yields
\begin{equation}\label{sy1}
    \Delta S_Y-\Delta I+\Delta S^{\rm (meas)}_{\rm env}\geq 0.
\end{equation}
Here, $\Delta I$ is the mutual information developed in the measurement. Notice however that the measurement in this description is any interaction between $X$ and $Y$ that creates correlations between the two systems, keeping $X$ constant.
During feedback, the demon does not change, hence $\dot S(Y(t))=0$. If the feedback process is cyclic, $\Delta S_X=0$, and if all the correlations created during the measurement are destroyed, then 
\begin{equation}\label{sy2}
    \Delta I+\Delta S^{\rm (fb)}_{\rm env}\geq 0.
\end{equation}
Finally, during the demon resetting:
\begin{equation}\label{sy3}
    -\Delta S_Y+\Delta S^{\rm (reset)}_{\rm env}\geq 0.
\end{equation}
The sum of the three inequalities (\ref{sy1}-\ref{sy3}) yields the standard second law for the total entropy production
\begin{equation}
    S_{\rm prod}=\Delta S^{\rm (meas)}_{\rm env}+\Delta S^{\rm (fb)}_{\rm env}+\Delta S^{\rm (reset)}_{\rm env}\geq 0.
\end{equation}
Therefore, Eq.~\eqref{totalproddt} comprises all our discussion on the Szil\'ard engine but also extends to generic dynamics where two systems evolve in time creating and destroying correlations \cite{Horowitz2014}.

\section{Information flows}
\label{sec:flows}

The previous discussion holds for an arbitrary evolution. However, for \eqref{totalproddt} to be meaningful, the evolution must be driven by an external agent 
(without feedback), which switches on and off the coupling between the system and the demon in the measurement and resets the demon to its initial state. Otherwise, the global system reaches a steady state where all terms in Eq.~\eqref{totalproddt} vanish in average.
Still, it is interesting to interpret as information engines autonomous motors that work in a non-equilibrium  stationary state (NESS). For this purpose, it is useful to introduce the notion of {\em information flows} \cite{Allahverdyan2009,horowitz_multipartite_2015,Horowitz2014}:
\begin{equation}
\label{Ixt}
\dot I_X(t)=\lim_{\tau\to 0^+}\frac{I(X(t+\tau);Y(t))-I(X(t);Y(t))}{\tau} 
\end{equation}
and
\begin{equation}
\label{Iyt}
\dot I_Y(t)=\lim_{\tau\to 0^+}\frac{I(X(t);Y(t+\tau))-I(X(t);Y(t))}{\tau} .
\end{equation}
In the NESS, $\dot I_X(t)+\dot I_Y(t)=\dot I(X(t);Y(t))=0$, and $\dot I_{X}(t)=-\dot I_{Y}(t)$. This is why these quantities  can be considered flows of a conserved quantity, although this interpretation is valid only in the stationary regime. 

If $ \dot I_{X}(t)$
is positive then the motion of $X$, with $Y$  fixed, increases the correlation between the two systems. Accordingly to our discussion in the previous section, we can interpret it as $X$ measuring $Y$. On the other hand, if $\dot I_X(t)<0$, the motion of system $X$ destroys correlations, like a demon using information to complete a feedback process and converting the free energy stored in correlations into work.

When  each system is coupled to  separate baths,
one can derive a second law for the entropy production in the stationary regime due to the dynamics of each system, $X$ or $Y$:
\begin{equation}\label{2ndflows}
    \begin{split}
        -\dot I_X+\dot S_{\rm env}^X &\geq& 0  \\
\dot I_X+\dot S_{\rm env}^Y &\geq& 0  
    \end{split}
\end{equation}
where  $\dot S_{\rm env}^{X,Y}$ is the entropy increase in the reservoir connected to system $X$ and $Y$, respectively. These partial second laws have been derived for  multipartite continuous and discrete Markovian systems \cite{Horowitz2014,horowitz_multipartite_2015} and they allow us to analyze the energetics of autonomous motors from the point of view of information flows. 
In Ref.~\cite{Horowitz2014}, Horowitz and Esposito apply it to two coupled quantum dots in contact with reservoirs with different temperatures and/or chemical potentials, a device that can operate as a pump or a refrigerator. Information flows and Eq.~\eqref{2ndflows} do not only provide an interesting interpretation of one of the dots acting as a Maxwell demon on the other, but are also useful to derive limitations to the energetics that go beyond the application of the standard second law to the whole device. It is possible as well  to evaluate separately the efficiency of the exchange between entropy and information in each dot \cite{Horowitz2014}. See also \cite{rosinberg2016} for an example of continuous classical degrees of freedom and \cite{ptaszynski2019b} for an extension to quantum systems.

Information flows are related to other magnitudes 
 that quantify the exchange of information among different systems.
In the context of  data  analysis, Schreiber introduced  the so-called {\em transfer entropy} \cite{schreiber2000} to find causal relationships between two time series, whereas  Liang and Kleeman \cite{liang2005} studied similar information transfers in dynamical systems. Although transfer entropies have been criticized as a tool to find causal relations \cite{james2016}, they have been proven useful to derive fluctuation theorems in bipartite systems \cite{Hartich2014} and  general causal networks \cite{Ito2013}. The relationship between transfer entropies and information flows has been studied in detail in \cite{Allahverdyan2009,cafaro2016,horowitz_multipartite_2015,Horowitz2014b,kiwata2022}.

\section{Conclusions: what is information?}
\label{sec:conclusions}

Thermodynamics of information makes a precise assessment of the effect of information on the second law and clarifies the thermodynamic cost of information operations. The key concept is the mutual information between the state of a system and the outcome of a measurement since, as shown in \eqref{mituali2}, it is equal to the decrease of Shannon entropy  due to the measurement, but also to the difference, due to correlations, between the entropy of a composite system and the sum of the entropies of its parts. To relate  Shannon entropy with the energetics of a process, we have introduced in section \ref{sec:shantherm} the loss of information that one would expect when a system is in contact with reservoirs at equilibrium. This analysis yields the definition of another relevant concept: the non-equilibrium free energy \eqref{noneqfe}, which determines the minimal work necessary to complete an isothermal process connecting non-equilibrium states.

We have shown that the states relevant to information processing are out of equilibrium, either because they are the result of a Bayesian update, like \eqref{bayesupdate}, or because they describe memories with arbitrary probabilistic informational states, as in \eqref{rhoinfo}. In both cases, non-equilibrium free energy establishes limitations to the energetics of isothermal processes involving information. These limitations are extensions of the second law to informational non-equilibrium states. In this framework, information appears as a novel thermodynamic resource and those extensions of the second law are indispensable to assess the efficiency of its exploitation.

We have also used non-equilibrium free energy to prove that the Szil\'ard engine and any  isothermal feedback process are compatible with the standard second law of thermodynamics. For this purpose, we have  considered the physical nature of the demon and, in doing so, we not only solve the fundamental question posed by Maxwell more than one century ago, but  open as well the possibility of interpreting certain devices like motors, refrigerators, and other chemical or thermal machines, as transducers that convert information into work and vice-versa. For this interpretation to hold, thermal fluctuations and correlations must be essential ingredients in the functioning of these machines. This is why information thermodynamics is closely related to ---or can even be considered as a branch of---  a recent field of research, stochastic thermodynamics, which studies the energetics of processes at the micro- and nano-scale, incorporating fluctuations \cite{Peliti,Sekimoto}.

In this article, we have restricted ourselves mostly to classical systems. The reason is that, in an isothermal process, a quantum system  rapidly adopts a state which is diagonal in the eigenbasis of the Hamiltonian. In these scenarios, coherences are lost and the system can be studied with the same tools as those developed for classical systems. In fact, most of the concepts that we have introduced in the previous sections can be straightforwardly extended to quantum systems. Nevertheless, in the last years, there has been an increasing interest in phenomena that combine in a nontrivial way quantum and thermal effects \cite{Anders2017}. Zurek \cite{Zurek1984} and Lloyd \cite{Lloyd:1997ha} already found  distinctive aspects of the quantum Szil\'ard engine  and Kim {\em et al} explored the effect of quantum statistics in multiparticle Szil\'ard cycles  with bosons and fermions \cite{Kim2011b}. The mutual information between quantum systems can be split into a classical and a purely quantum part due to entanglement, known as {\em  quantum discord}, and whose role in quantum Maxwell demons have been analyzed in \cite{Park2013,Zurek2003}. Finally, some setups involving the coupling between two quantum states can be interpreted as Maxwell demons that use information. This is the case of the  experiment  reported in  \cite{cottet2017} where the coupling between a transmon qubit and the radiation in a microwave cavity can be used to extract work.

From a more fundamental point of view, the interplay between information and matter is nowadays considered a crucial issue for our understanding of the physical world. In 1992, Wheeler summarized in  the celebrated lemma ``it from bit'' \cite{wheeler}  the idea  that information is the basic concept upon which every physical theory should be built. Since then, a number of physicists and philosophers have attempted to follow this research program. Thermodynamics of information, on the other hand, pursues more modest goals, but it does provide some insight into  how information is implemented in physical systems. In this article, we have seen two basic aspects of the physical nature of information. The first one is the characteristics  of the informational states
introduced in section \ref{sec:landauer}, namely, long lifetimes and interchangeability. The second one is  the interpretation of information in feedback engines as
an exchange between work and the free energy $TI(X;Y)$ stored in the correlations between two systems, $X$ and $Y$. Each of these two aspects has yielded interesting lines of research. The first one explores the energetics of processes in systems where there is a huge separation of time scales \cite{Roldan2014}, whereas the second one has inspired the notion of information flows in autonomous systems \cite{Horowitz2014}. We believe that any reflection about information as a fundamental concept in physics should account for and/or utilize these two aspects.

\section*{Acknowledgements}

This article summarizes work done over more than ten years and it would not have been possible without the interaction with many colleagues, friends, and students from different summer schools. 
A special mention deserve Jordan Horowitz, with whom I have done a great part of my contributions to the field, and Takahiro Sagawa, who has been one of the most active and influencing researchers in the recent application of stochastic thermodynamics to information processing. Together with Jordan and Takahiro, we wrote a review article \cite{Parrondo2015}, which shaped most of my ideas on the thermodynamics of information. The preparation of this review has been financially supported by the Spanish Government (Grant {\sc FLUID}. Ref: PID2020-113455GB-I00) and the Foundational Questions Institute  (FQXi) under the program ``Information as Fuel'' (Grant {\sc NanoQit}. Ref FQXi-IAF19-01). I appreciate the hospitality of Natalia Ares at University of Oxford and  discussions with the whole {\sc NanoQit} team on the content and organization of this article. Finally, I am indebted to Jordan Horowitz, Takahiro Sagawa, and Massimiliano Esposito for their careful reading of the article and their helpful suggestions.





%
%





\bibliographystyle{cas-model2-names}

\bibliography{book.bib}

\begin{thebibliography}{69}
\expandafter\ifx\csname natexlab\endcsname\relax\def\natexlab#1{#1}\fi
\providecommand{\url}[1]{\texttt{#1}}
\providecommand{\href}[2]{#2}
\providecommand{\path}[1]{#1}
\providecommand{\DOIprefix}{doi:}
\providecommand{\ArXivprefix}{arXiv:}
\providecommand{\URLprefix}{URL: }
\providecommand{\Pubmedprefix}{pmid:}
\providecommand{\doi}[1]{\href{http://dx.doi.org/#1}{\path{#1}}}
\providecommand{\Pubmed}[1]{\href{pmid:#1}{\path{#1}}}
\providecommand{\bibinfo}[2]{#2}
\ifx\xfnm\relax \def\xfnm[#1]{\unskip,\space#1}\fi
\bibitem[{Allahverdyan et~al.(2009)Allahverdyan, Janzing and
  Mahler}]{Allahverdyan2009}
\bibinfo{author}{Allahverdyan, A.E.}, \bibinfo{author}{Janzing, D.},
  \bibinfo{author}{Mahler, G.}, \bibinfo{year}{2009}.
\newblock \bibinfo{title}{Thermodynamic efficiency of information and heat
  flow}.
\newblock \bibinfo{journal}{Journal of Statistical Mechanics: Theory and
  Experiment} \bibinfo{volume}{2009}, \bibinfo{pages}{P09011}.
\newblock \DOIprefix\doi{10.1088/1742-5468/2009/09/P09011}.
  \bibinfo{note}{publisher: IOP Publishing}.
\bibitem[{Allahverdyan and Saakian(2008)}]{Allahverdyan2008}
\bibinfo{author}{Allahverdyan, A.E.}, \bibinfo{author}{Saakian, D.B.},
  \bibinfo{year}{2008}.
\newblock \bibinfo{title}{Thermodynamics of adiabatic feedback control}.
\newblock \bibinfo{journal}{Europhys. Lett.} \bibinfo{volume}{81},
  \bibinfo{pages}{30003}.
\bibitem[{Anders and Esposito(2017)}]{Anders2017}
\bibinfo{author}{Anders, J.}, \bibinfo{author}{Esposito, M.},
  \bibinfo{year}{2017}.
\newblock \bibinfo{title}{Focus on quantum thermodynamics}.
\newblock \bibinfo{journal}{New Journal of Physics} \bibinfo{volume}{19},
  \bibinfo{pages}{010201}.
\newblock \DOIprefix\doi{10.1088/1367-2630/19/1/010201}.
\bibitem[{Barato and Seifert(2014)}]{Barato2014b}
\bibinfo{author}{Barato, A.}, \bibinfo{author}{Seifert, U.},
  \bibinfo{year}{2014}.
\newblock \bibinfo{title}{Stochastic thermodynamics with information
  reservoirs}.
\newblock \bibinfo{journal}{Phys. Rev. E} \bibinfo{volume}{90},
  \bibinfo{pages}{042150}.
\bibitem[{Bennett(1982)}]{Bennett1982b}
\bibinfo{author}{Bennett, C.}, \bibinfo{year}{1982}.
\newblock \bibinfo{title}{{The thermodynamics of computation---a review}}.
\newblock \bibinfo{journal}{Int. J. Theor. Phys.} \bibinfo{volume}{21},
  \bibinfo{pages}{905--940}.
\newblock \bibinfo{note}{Reprinted in \cite{Leff}}.
\bibitem[{Berut et~al.(2011)Berut, Arakelyan, Petrosyan, Ciliberto,
  Dillenschneider and Lutz}]{Berut2011}
\bibinfo{author}{Berut, A.}, \bibinfo{author}{Arakelyan, A.},
  \bibinfo{author}{Petrosyan, A.}, \bibinfo{author}{Ciliberto, S.},
  \bibinfo{author}{Dillenschneider, R.}, \bibinfo{author}{Lutz, E.},
  \bibinfo{year}{2011}.
\newblock \bibinfo{title}{Experimental verifciation of landauer's principle
  linking information and thermodynamics}.
\newblock \bibinfo{journal}{Nature} \bibinfo{volume}{483},
  \bibinfo{pages}{187--189}.
\bibitem[{Bustamante et~al.(2001)Bustamante, Keller and Oster}]{Bustamante2001}
\bibinfo{author}{Bustamante, C.}, \bibinfo{author}{Keller, D.},
  \bibinfo{author}{Oster, G.}, \bibinfo{year}{2001}.
\newblock \bibinfo{title}{{The Physics of Molecular Motors}}.
\newblock \bibinfo{journal}{Acc. Chem. Res.} \bibinfo{volume}{34},
  \bibinfo{pages}{412--420}.
\bibitem[{Cafaro et~al.(2016)Cafaro, Ali and Giffin}]{cafaro2016}
\bibinfo{author}{Cafaro, C.}, \bibinfo{author}{Ali, S.A.},
  \bibinfo{author}{Giffin, A.}, \bibinfo{year}{2016}.
\newblock \bibinfo{title}{Thermodynamic aspects of information transfer in
  complex dynamical systems}.
\newblock \bibinfo{journal}{Physical Review E} \bibinfo{volume}{93},
  \bibinfo{pages}{022114}.
\newblock \DOIprefix\doi{10.1103/PhysRevE.93.022114}. \bibinfo{note}{publisher:
  American Physical Society}.
\bibitem[{Cottet et~al.(2017)Cottet, Jezouin, Bretheau, Campagne-Ibarcq,
  Ficheux, Anders, Auff\`eves, Azouit, Rouchon and Huard}]{cottet2017}
\bibinfo{author}{Cottet, N.}, \bibinfo{author}{Jezouin, S.},
  \bibinfo{author}{Bretheau, L.}, \bibinfo{author}{Campagne-Ibarcq, P.},
  \bibinfo{author}{Ficheux, Q.}, \bibinfo{author}{Anders, J.},
  \bibinfo{author}{Auff\`eves, A.}, \bibinfo{author}{Azouit, R.},
  \bibinfo{author}{Rouchon, P.}, \bibinfo{author}{Huard, B.},
  \bibinfo{year}{2017}.
\newblock \bibinfo{title}{Observing a quantum {Maxwell} demon at work}.
\newblock \bibinfo{journal}{Proceedings of the National Academy of Sciences}
  \bibinfo{volume}{114}, \bibinfo{pages}{7561--7564}.
\newblock \DOIprefix\doi{10.1073/pnas.1704827114}. \bibinfo{note}{publisher:
  Proceedings of the National Academy of Sciences}.
\bibitem[{Cover and Thomas(2006)}]{Cover}
\bibinfo{author}{Cover, T.M.}, \bibinfo{author}{Thomas, J.A.},
  \bibinfo{year}{2006}.
\newblock \bibinfo{title}{Elements of Information Theory}.
\newblock \bibinfo{edition}{Second} ed.,
  \bibinfo{publisher}{Wiley-Interscience, New York, NY}.
\bibitem[{Ehrich et~al.(2019)Ehrich, Esposito, Barra and Parrondo}]{Ehrich2019}
\bibinfo{author}{Ehrich, J.}, \bibinfo{author}{Esposito, M.},
  \bibinfo{author}{Barra, F.}, \bibinfo{author}{Parrondo, J.M.},
  \bibinfo{year}{2019}.
\newblock \bibinfo{title}{Micro-reversibility and thermalization with
  collisional baths}.
\newblock \bibinfo{journal}{Physica A: Statistical Mechanics and its
  Applications} ,
  \bibinfo{pages}{122108}\DOIprefix\doi{https://doi.org/10.1016/j.physa.2019.122108}.
\bibitem[{Esposito and Van~den Broeck(2011)}]{Esposito2011}
\bibinfo{author}{Esposito, M.}, \bibinfo{author}{Van~den Broeck, C.},
  \bibinfo{year}{2011}.
\newblock \bibinfo{title}{Second law and landauer principle far from
  equilibrium}.
\newblock \bibinfo{journal}{Europhys. Lett.} \bibinfo{volume}{95},
  \bibinfo{pages}{40004}.
\bibitem[{Esposito et~al.(2010)Esposito, Lindenberg and den
  Broeck}]{Esposito2009}
\bibinfo{author}{Esposito, M.}, \bibinfo{author}{Lindenberg, K.},
  \bibinfo{author}{den Broeck, C.V.}, \bibinfo{year}{2010}.
\newblock \bibinfo{title}{Entropy production as correlation between system and
  reservoir}.
\newblock \bibinfo{journal}{New Journal of Physics} \bibinfo{volume}{12},
  \bibinfo{pages}{013013}.
\newblock \DOIprefix\doi{10.1088/1367-2630/12/1/013013}.
\bibitem[{Feng et~al.(2021)Feng, Ovalle, Seale, Lee, Kim, Astumian and
  Stoddart}]{Feng2021}
\bibinfo{author}{Feng, Y.}, \bibinfo{author}{Ovalle, M.},
  \bibinfo{author}{Seale, J.S.W.}, \bibinfo{author}{Lee, C.K.},
  \bibinfo{author}{Kim, D.J.}, \bibinfo{author}{Astumian, R.D.},
  \bibinfo{author}{Stoddart, J.F.}, \bibinfo{year}{2021}.
\newblock \bibinfo{title}{{Molecular Pumps and Motors}}.
\newblock \bibinfo{journal}{J. Am. Chem. Soc.} \bibinfo{volume}{143},
  \bibinfo{pages}{5569--5591}.
\bibitem[{Feynman et~al.(1966)Feynman, Leighton and Sands}]{Feynman}
\bibinfo{author}{Feynman, R.P.}, \bibinfo{author}{Leighton, R.B.},
  \bibinfo{author}{Sands, M.}, \bibinfo{year}{1966}.
\newblock \bibinfo{title}{The Feynman Lectures on Physics. Vol I}.
\newblock \bibinfo{publisher}{Addison-Wesley}.
\newblock \bibinfo{note}{Chap. 46}.
\bibitem[{Hartich et~al.(2014)Hartich, Barato and Seifert}]{Hartich2014}
\bibinfo{author}{Hartich, D.}, \bibinfo{author}{Barato, A.C.},
  \bibinfo{author}{Seifert, U.}, \bibinfo{year}{2014}.
\newblock \bibinfo{title}{Stochastic thermodynamics of bipartite systems:
  transfer entropy inequalities and a maxwell's demon interpretation}.
\newblock \bibinfo{journal}{J. Stat. Mech.: Theor. Exp.} ,
  \bibinfo{pages}{P02016}.
\bibitem[{Horowitz(2015)}]{horowitz_multipartite_2015}
\bibinfo{author}{Horowitz, J.M.}, \bibinfo{year}{2015}.
\newblock \bibinfo{title}{Multipartite information flow for multiple {Maxwell}
  demons}.
\newblock \bibinfo{journal}{Journal of Statistical Mechanics: Theory and
  Experiment} \bibinfo{volume}{2015}, \bibinfo{pages}{P03006}.
\newblock \DOIprefix\doi{10.1088/1742-5468/2015/03/P03006}.
  \bibinfo{note}{publisher: IOP Publishing}.
\bibitem[{Horowitz and Esposito(2014)}]{Horowitz2014}
\bibinfo{author}{Horowitz, J.M.}, \bibinfo{author}{Esposito, M.},
  \bibinfo{year}{2014}.
\newblock \bibinfo{title}{Thermodynamics with continuous information flow}.
\newblock \bibinfo{journal}{Phys. Rev. X} \bibinfo{volume}{4},
  \bibinfo{pages}{031015}.
\bibitem[{Horowitz and Parrondo(2011a)}]{Horowitz2011b}
\bibinfo{author}{Horowitz, J.M.}, \bibinfo{author}{Parrondo, J.M.R.},
  \bibinfo{year}{2011}a.
\newblock \bibinfo{title}{Designing optimal discrete-feedback thermodynamic
  engines}.
\newblock \bibinfo{journal}{New J. Phys.} \bibinfo{volume}{13},
  \bibinfo{pages}{123019}.
\bibitem[{Horowitz and Parrondo(2011b)}]{Horowitz2011}
\bibinfo{author}{Horowitz, J.M.}, \bibinfo{author}{Parrondo, J.M.R.},
  \bibinfo{year}{2011}b.
\newblock \bibinfo{title}{Thermodynamic reversibility in feedback processes}.
\newblock \bibinfo{journal}{Europhys. Lett.} \bibinfo{volume}{95},
  \bibinfo{pages}{10005}.
\bibitem[{Horowitz et~al.(2013)Horowitz, Sagawa and Parrondo}]{Horowitz2013}
\bibinfo{author}{Horowitz, J.M.}, \bibinfo{author}{Sagawa, T.},
  \bibinfo{author}{Parrondo, J.M.R.}, \bibinfo{year}{2013}.
\newblock \bibinfo{title}{Imitating chemical motors with optimal information
  motors}.
\newblock \bibinfo{journal}{Physical Review Letters} \bibinfo{volume}{111},
  \bibinfo{pages}{010602}.
\bibitem[{Horowitz and Sandberg(2014)}]{horowitz_second-law-like_2014}
\bibinfo{author}{Horowitz, J.M.}, \bibinfo{author}{Sandberg, H.},
  \bibinfo{year}{2014}.
\newblock \bibinfo{title}{Second-law-like inequalities with information and
  their interpretations}.
\newblock \bibinfo{journal}{New Journal of Physics} \bibinfo{volume}{16},
  \bibinfo{pages}{125007}.
\newblock \DOIprefix\doi{10.1088/1367-2630/16/12/125007}.
  \bibinfo{note}{publisher: IOP Publishing}.
\bibitem[{Horowitz and Vaikuntanathan(2010)}]{Horowitz2010}
\bibinfo{author}{Horowitz, J.M.}, \bibinfo{author}{Vaikuntanathan, S.},
  \bibinfo{year}{2010}.
\newblock \bibinfo{title}{Nonequilibrium detailed fluctuation theorem for
  discrete feedback}.
\newblock \bibinfo{journal}{Phys. Rev. E} \bibinfo{volume}{82},
  \bibinfo{pages}{061120}.
\bibitem[{Ito and Sagawa(2013)}]{Ito2013}
\bibinfo{author}{Ito, S.}, \bibinfo{author}{Sagawa, T.}, \bibinfo{year}{2013}.
\newblock \bibinfo{title}{Information thermodynamics on causal networks}.
\newblock \bibinfo{journal}{Physical Review Letters} \bibinfo{volume}{111},
  \bibinfo{pages}{180603}.
\bibitem[{James et~al.(2016)James, Barnett and Crutchfield}]{james2016}
\bibinfo{author}{James, R.G.}, \bibinfo{author}{Barnett, N.},
  \bibinfo{author}{Crutchfield, J.P.}, \bibinfo{year}{2016}.
\newblock \bibinfo{title}{Information {Flows}? {A} {Critique} of {Transfer}
  {Entropies}}.
\newblock \bibinfo{journal}{Physical Review Letters} \bibinfo{volume}{116},
  \bibinfo{pages}{238701}.
\newblock \DOIprefix\doi{10.1103/PhysRevLett.116.238701}.
  \bibinfo{note}{publisher: American Physical Society}.
\bibitem[{Kim et~al.(2011)Kim, Sagawa, De~Liberato and Ueda}]{Kim2011b}
\bibinfo{author}{Kim, S.W.}, \bibinfo{author}{Sagawa, T.},
  \bibinfo{author}{De~Liberato, S.}, \bibinfo{author}{Ueda, M.},
  \bibinfo{year}{2011}.
\newblock \bibinfo{title}{Quantum {S}zilard engine}.
\newblock \bibinfo{journal}{Physical Review Letters} \bibinfo{volume}{106},
  \bibinfo{pages}{070401}.
\bibitem[{Kiwata(2022)}]{kiwata2022}
\bibinfo{author}{Kiwata, H.}, \bibinfo{year}{2022}.
\newblock \bibinfo{title}{Relationship between {Schreiber}'s transfer entropy
  and {Liang}-{Kleeman} information flow from the perspective of stochastic
  thermodynamics}.
\newblock \bibinfo{journal}{Physical Review E} \bibinfo{volume}{105},
  \bibinfo{pages}{044130}.
\newblock \DOIprefix\doi{10.1103/PhysRevE.105.044130}.
  \bibinfo{note}{publisher: American Physical Society}.
\bibitem[{Koski et~al.(2014)Koski, Maisi, Pekola and Averin}]{Koski13786}
\bibinfo{author}{Koski, J.V.}, \bibinfo{author}{Maisi, V.F.},
  \bibinfo{author}{Pekola, J.P.}, \bibinfo{author}{Averin, D.V.},
  \bibinfo{year}{2014}.
\newblock \bibinfo{title}{Experimental realization of a szilard engine with a
  single electron}.
\newblock \bibinfo{journal}{Proceedings of the National Academy of Sciences}
  \bibinfo{volume}{111}, \bibinfo{pages}{13786--13789}.
\newblock \DOIprefix\doi{10.1073/pnas.1406966111}.
\bibitem[{{Landauer}(1961)}]{Landauer1961}
\bibinfo{author}{{Landauer}, R.}, \bibinfo{year}{1961}.
\newblock \bibinfo{title}{Irreversibility and heat generation in the computing
  process}.
\newblock \bibinfo{journal}{IBM Journal of Research and Development}
  \bibinfo{volume}{5}, \bibinfo{pages}{183--191}.
\newblock \DOIprefix\doi{10.1147/rd.53.0183}.
\bibitem[{Leff and Rex(1990)}]{Leff}
\bibinfo{editor}{Leff, H.S.}, \bibinfo{editor}{Rex, A.F.} (Eds.),
  \bibinfo{year}{1990}.
\newblock \bibinfo{title}{Maxwell's Demon: Entropy, Information, Computing}.
\newblock \bibinfo{publisher}{Princeton University Press, New Jersey}.
\bibitem[{Liang and Kleeman(2005)}]{liang2005}
\bibinfo{author}{Liang, X.S.}, \bibinfo{author}{Kleeman, R.},
  \bibinfo{year}{2005}.
\newblock \bibinfo{title}{Information {Transfer} between {Dynamical} {System}
  {Components}}.
\newblock \bibinfo{journal}{Physical Review Letters} \bibinfo{volume}{95},
  \bibinfo{pages}{244101}.
\newblock \DOIprefix\doi{10.1103/PhysRevLett.95.244101}.
  \bibinfo{note}{publisher: American Physical Society}.
\bibitem[{Lloyd(1989)}]{Lloyd1989}
\bibinfo{author}{Lloyd, S.}, \bibinfo{year}{1989}.
\newblock \bibinfo{title}{Use of mutual information to decrease entropy:
  Implications for the second law of thermodynamics}.
\newblock \bibinfo{journal}{Phys. Rev. A} \bibinfo{volume}{39},
  \bibinfo{pages}{5378--5386}.
\bibitem[{Lloyd(1997)}]{Lloyd:1997ha}
\bibinfo{author}{Lloyd, S.}, \bibinfo{year}{1997}.
\newblock \bibinfo{title}{{Quantum-mechanical {M}axwell's demon}}.
\newblock \bibinfo{journal}{Phys. Rev. A} \bibinfo{volume}{56},
  \bibinfo{pages}{3374--3382}.
\bibitem[{Mandal and Jarzynski(2012)}]{Mandal2012}
\bibinfo{author}{Mandal, D.}, \bibinfo{author}{Jarzynski, C.},
  \bibinfo{year}{2012}.
\newblock \bibinfo{title}{Work and information processing in a solvable model
  of maxwell's demon}.
\newblock \bibinfo{journal}{Proc. Nat. Ac. Sci.} .
\bibitem[{Ptaszy\ifmmode~\acute{n}\else \'{n}\fi{}ski and
  Esposito(2019)}]{ptaszynski2019b}
\bibinfo{author}{Ptaszy\ifmmode~\acute{n}\else \'{n}\fi{}ski, K.},
  \bibinfo{author}{Esposito, M.}, \bibinfo{year}{2019}.
\newblock \bibinfo{title}{Thermodynamics of quantum information flows}.
\newblock \bibinfo{journal}{Phys. Rev. Lett.} \bibinfo{volume}{122},
  \bibinfo{pages}{150603}.
\newblock \DOIprefix\doi{10.1103/PhysRevLett.122.150603}.
\bibitem[{Park et~al.(2013)Park, Kim, Sagawa and Kim}]{Park2013}
\bibinfo{author}{Park, J.J.}, \bibinfo{author}{Kim, K.H.},
  \bibinfo{author}{Sagawa, T.}, \bibinfo{author}{Kim, S.W.},
  \bibinfo{year}{2013}.
\newblock \bibinfo{title}{Heat engine driven by purely quantum information}.
\newblock \bibinfo{journal}{Physical Review Letters} \bibinfo{volume}{111},
  \bibinfo{pages}{230402}.
\bibitem[{Parrondo(2001)}]{Parrondo2001}
\bibinfo{author}{Parrondo, J.M.R.}, \bibinfo{year}{2001}.
\newblock \bibinfo{title}{The {S}zilard engine revisited: Entropy, macroscopic
  randomness, and symmetry breaking phase transitions}.
\newblock \bibinfo{journal}{Chaos} \bibinfo{volume}{11},
  \bibinfo{pages}{725--733}.
\bibitem[{Parrondo and Espa{\~n}ol(1996)}]{Parrondo1996}
\bibinfo{author}{Parrondo, J.M.R.}, \bibinfo{author}{Espa{\~n}ol, P.},
  \bibinfo{year}{1996}.
\newblock \bibinfo{title}{Criticism of {F}eynman's analysis of the ratchet as
  an engine}.
\newblock \bibinfo{journal}{American Journal of Physics} \bibinfo{volume}{64},
  \bibinfo{pages}{1125--1130}.
\newblock \DOIprefix\doi{10.1119/1.18393}.
\bibitem[{Parrondo et~al.(2015)Parrondo, Horowitz and Sagawa}]{Parrondo2015}
\bibinfo{author}{Parrondo, J.M.R.}, \bibinfo{author}{Horowitz, J.M.},
  \bibinfo{author}{Sagawa, T.}, \bibinfo{year}{2015}.
\newblock \bibinfo{title}{Thermodynamics of information}.
\newblock \bibinfo{journal}{Nature Physics} \bibinfo{volume}{11},
  \bibinfo{pages}{131--139}.
\newblock \DOIprefix\doi{10.1038/nphys3230}.
\bibitem[{Pekola and Khaymovich(2019)}]{pekola2019}
\bibinfo{author}{Pekola, J.}, \bibinfo{author}{Khaymovich, I.},
  \bibinfo{year}{2019}.
\newblock \bibinfo{title}{Thermodynamics in {Single}-{Electron} {Circuits} and
  {Superconducting} {Qubits}}.
\newblock \bibinfo{journal}{Annual Review of Condensed Matter Physics}
  \bibinfo{volume}{10}, \bibinfo{pages}{193--212}.
\newblock \DOIprefix\doi{10.1146/annurev-conmatphys-033117-054120}.
  \bibinfo{note}{\_eprint:
  https://doi.org/10.1146/annurev-conmatphys-033117-054120}.
\bibitem[{Peliti and Pigolotti(2021)}]{Peliti}
\bibinfo{author}{Peliti, L.}, \bibinfo{author}{Pigolotti, S.},
  \bibinfo{year}{2021}.
\newblock \bibinfo{title}{Stochastic Thermodynamics}.
\newblock \bibinfo{publisher}{Princeton University Press}.
\bibitem[{Ponmurugan(2010)}]{Ponmurugan2010}
\bibinfo{author}{Ponmurugan, M.}, \bibinfo{year}{2010}.
\newblock \bibinfo{title}{Generalized detailed fluctuation theorem under
  nonequilibrium feedback control}.
\newblock \bibinfo{journal}{Phys. Rev. E} \bibinfo{volume}{82},
  \bibinfo{pages}{031129}.
\newblock \DOIprefix\doi{10.1103/PhysRevE.82.031129}.
\bibitem[{Price(1982)}]{oral_historyclaude_2021}
\bibinfo{author}{Price, R.}, \bibinfo{year}{1982}.
\newblock \bibinfo{title}{Oral-{History}:{Claude} {E}. {Shannon}}.
\newblock \bibinfo{journal}{Engineering and Technology History Wiki (ETHW)}
  \URLprefix \url{https://ethw.org/Oral-History:Claude_E._Shannon}.
\bibitem[{Proesmans et~al.(2020)Proesmans, Ehrich and
  Bechhoefer}]{Proesmans2020}
\bibinfo{author}{Proesmans, K.}, \bibinfo{author}{Ehrich, J.},
  \bibinfo{author}{Bechhoefer, J.}, \bibinfo{year}{2020}.
\newblock \bibinfo{title}{Finite-time landauer principle}.
\newblock \bibinfo{journal}{Phys. Rev. Lett.} \bibinfo{volume}{125},
  \bibinfo{pages}{100602}.
\newblock \DOIprefix\doi{10.1103/PhysRevLett.125.100602}.
\bibitem[{Ptaszy{\'{n}}ski and Esposito(2019)}]{Ptaszynski2019}
\bibinfo{author}{Ptaszy{\'{n}}ski, K.}, \bibinfo{author}{Esposito, M.},
  \bibinfo{year}{2019}.
\newblock \bibinfo{title}{{Entropy Production in Open Systems: The Predominant
  Role of Intraenvironment Correlations}}.
\newblock \bibinfo{journal}{Physical Review Letters} \bibinfo{volume}{123},
  \bibinfo{pages}{200603}.
\bibitem[{Reimann(2002)}]{Reimann:2002hs}
\bibinfo{author}{Reimann, P.}, \bibinfo{year}{2002}.
\newblock \bibinfo{title}{{Brownian motors: noisy transport far from
  equilibrium}}.
\newblock \bibinfo{journal}{Phys. Rep.} \bibinfo{volume}{361},
  \bibinfo{pages}{57--265}.
\bibitem[{Rold\'an et~al.(2014)Rold\'an, Mart\'{\i}nez, Parrondo and
  Petrov}]{Roldan2014}
\bibinfo{author}{Rold\'an, E.}, \bibinfo{author}{Mart\'{\i}nez, I.A.},
  \bibinfo{author}{Parrondo, J.M.R.}, \bibinfo{author}{Petrov, D.},
  \bibinfo{year}{2014}.
\newblock \bibinfo{title}{Universal features in the energetics of symmetry
  breaking}.
\newblock \bibinfo{journal}{Nature Physics} .
\bibitem[{Rosinberg and Horowitz(2016)}]{rosinberg2016}
\bibinfo{author}{Rosinberg, M.L.}, \bibinfo{author}{Horowitz, J.M.},
  \bibinfo{year}{2016}.
\newblock \bibinfo{title}{Continuous information flow fluctuations}.
\newblock \bibinfo{journal}{EPL (Europhysics Letters)} \bibinfo{volume}{116},
  \bibinfo{pages}{10007}.
\newblock \DOIprefix\doi{10.1209/0295-5075/116/10007}.
  \bibinfo{note}{publisher: IOP Publishing}.
\bibitem[{Sagawa(2012a)}]{Sagawa2012b}
\bibinfo{author}{Sagawa, T.}, \bibinfo{year}{2012}a.
\newblock \bibinfo{title}{{Thermodynamics of Information Processing in Small
  Systems*)}}.
\newblock \bibinfo{journal}{Progress of Theoretical Physics}
  \bibinfo{volume}{127}, \bibinfo{pages}{1--56}.
\newblock \DOIprefix\doi{10.1143/PTP.127.1}.
\bibitem[{Sagawa(2012b)}]{sagawa2012thermodynamics}
\bibinfo{author}{Sagawa, T.}, \bibinfo{year}{2012}b.
\newblock \bibinfo{title}{{Thermodynamics of Information Processing in Small
  Systems}}.
\newblock Springer Theses, \bibinfo{publisher}{Springer}.
\bibitem[{Sagawa(2014)}]{Sagawa2014}
\bibinfo{author}{Sagawa, T.}, \bibinfo{year}{2014}.
\newblock \bibinfo{title}{Thermodynamic and logical reversibilities revisited}.
\newblock \bibinfo{journal}{Journal of Statistical Mechanics: Theory and
  Experiment} \bibinfo{volume}{2014}, \bibinfo{pages}{P03025}.
\newblock \DOIprefix\doi{10.1088/1742-5468/2014/03/p03025}.
\bibitem[{Sagawa and Ueda(2008)}]{Sagawa2008}
\bibinfo{author}{Sagawa, T.}, \bibinfo{author}{Ueda, M.}, \bibinfo{year}{2008}.
\newblock \bibinfo{title}{Second law of thermodynamics with discrete quantum
  feedback control}.
\newblock \bibinfo{journal}{Physical Review Letters} \bibinfo{volume}{100},
  \bibinfo{pages}{080403}.
\bibitem[{Sagawa and Ueda(2009)}]{Sagawa2009}
\bibinfo{author}{Sagawa, T.}, \bibinfo{author}{Ueda, M.}, \bibinfo{year}{2009}.
\newblock \bibinfo{title}{Minimal energy cost for thermodynamic information
  processing: Measurement and information erasure}.
\newblock \bibinfo{journal}{Physical Review Letters} \bibinfo{volume}{102},
  \bibinfo{pages}{250602}.
\bibitem[{Sagawa and Ueda(2010)}]{Sagawa2010}
\bibinfo{author}{Sagawa, T.}, \bibinfo{author}{Ueda, M.}, \bibinfo{year}{2010}.
\newblock \bibinfo{title}{Generalized jarzynski equality under nonequilibrium
  feedback control}.
\newblock \bibinfo{journal}{Physical Review Letters} \bibinfo{volume}{104},
  \bibinfo{pages}{090602}.
\bibitem[{Sagawa and Ueda(2012)}]{Sagawa2012}
\bibinfo{author}{Sagawa, T.}, \bibinfo{author}{Ueda, M.}, \bibinfo{year}{2012}.
\newblock \bibinfo{title}{Fluctuation theorem with information exchange: Role
  of correlations in stochastic thermodynamics}.
\newblock \bibinfo{journal}{Physical Review Letters} \bibinfo{volume}{109},
  \bibinfo{pages}{180602}.
\bibitem[{Saha et~al.(2021)Saha, Lucero, Ehrich, Sivak and
  Bechhoefer}]{Saha2021b}
\bibinfo{author}{Saha, T.K.}, \bibinfo{author}{Lucero, J.N.E.},
  \bibinfo{author}{Ehrich, J.}, \bibinfo{author}{Sivak, D.A.},
  \bibinfo{author}{Bechhoefer, J.}, \bibinfo{year}{2021}.
\newblock \bibinfo{title}{Maximizing power and velocity of an information
  engine}.
\newblock \bibinfo{journal}{Proceedings of the National Academy of Sciences}
  \bibinfo{volume}{118}, \bibinfo{pages}{e2023356118}.
\newblock \DOIprefix\doi{10.1073/pnas.2023356118}.
\bibitem[{Schmitt et~al.(2015)Schmitt, Parrondo, Linke and
  Johansson}]{Schmitt2015}
\bibinfo{author}{Schmitt, R.K.}, \bibinfo{author}{Parrondo, J.M.R.},
  \bibinfo{author}{Linke, H.}, \bibinfo{author}{Johansson, J.},
  \bibinfo{year}{2015}.
\newblock \bibinfo{title}{Molecular motor efficiency is maximized in the
  presence of both power-stroke and rectification through feedback}.
\newblock \bibinfo{journal}{New Journal of Physics} \bibinfo{volume}{17}.
\newblock \DOIprefix\doi{10.1088/1367-2630/17/6/065011}.
\bibitem[{Schreiber(2000)}]{schreiber2000}
\bibinfo{author}{Schreiber, T.}, \bibinfo{year}{2000}.
\newblock \bibinfo{title}{Measuring {Information} {Transfer}}.
\newblock \bibinfo{journal}{Physical Review Letters} \bibinfo{volume}{85},
  \bibinfo{pages}{461--464}.
\newblock \DOIprefix\doi{10.1103/PhysRevLett.85.461}. \bibinfo{note}{publisher:
  American Physical Society}.
\bibitem[{Schumacher and Westmoreland(2010)}]{schumacher2010}
\bibinfo{author}{Schumacher, B.}, \bibinfo{author}{Westmoreland, M.},
  \bibinfo{year}{2010}.
\newblock \bibinfo{title}{Quantum Processes Systems, and Information}.
\newblock \bibinfo{publisher}{Cambridge University Press},
  \bibinfo{address}{USA}.
\bibitem[{Sekimoto(2010)}]{Sekimoto}
\bibinfo{author}{Sekimoto, K.}, \bibinfo{year}{2010}.
\newblock \bibinfo{title}{Stochastic Energetics}. volume \bibinfo{volume}{799}
  of \textit{\bibinfo{series}{Lect. Notes Phys.}}
\newblock \bibinfo{publisher}{Springer, Berlin Heidelberg}.
\bibitem[{Shannon(1948)}]{shannon}
\bibinfo{author}{Shannon, C.E.}, \bibinfo{year}{1948}.
\newblock \bibinfo{title}{A mathematical theory of communication}.
\newblock \bibinfo{journal}{Bell System Technical Journal}
  \bibinfo{volume}{27}, \bibinfo{pages}{379--423}.
\newblock \DOIprefix\doi{https://doi.org/10.1002/j.1538-7305.1948.tb01338.x}.
\bibitem[{Smoluchowski(1912)}]{Smoluchowski:1912}
\bibinfo{author}{Smoluchowski, M.v.}, \bibinfo{year}{1912}.
\newblock \bibinfo{title}{{Experimentell nachweisbare, der {\"U}blichen
  {T}hermodynamik widersprechende {M}olekularphenomene}}.
\newblock \bibinfo{journal}{Phys. Zeitshur.} \bibinfo{volume}{13},
  \bibinfo{pages}{1069}.
\bibitem[{Szil\'ard(1929)}]{Szilard1929}
\bibinfo{author}{Szil\'ard, L.}, \bibinfo{year}{1929}.
\newblock \bibinfo{title}{{\"u}ber die entropieverminderung in einem
  thermodynamischen system bei eingriffen intelligenter wesen}.
\newblock \bibinfo{journal}{Zeitschrift f{\"u}r Physik} \bibinfo{volume}{53},
  \bibinfo{pages}{840--856}.
\newblock \DOIprefix\doi{10.1007/BF01341281}. \bibinfo{note}{{\em On the
  decrease of entropy in a thermodynamic system by the intervention of
  intelligent beings}. Translated into English and reprinted in \cite{Leff}}.
\bibitem[{Touchette and Lloyd(2000)}]{Touchette2000}
\bibinfo{author}{Touchette, H.}, \bibinfo{author}{Lloyd, S.},
  \bibinfo{year}{2000}.
\newblock \bibinfo{title}{Information-theoreitc limits of control}.
\newblock \bibinfo{journal}{Physical Review Letters} \bibinfo{volume}{84},
  \bibinfo{pages}{1156--1159}.
\bibitem[{Toyabe et~al.(2010)Toyabe, Sagawa, Ueda, Muneyuki and
  Sano}]{Toyabe2010}
\bibinfo{author}{Toyabe, S.}, \bibinfo{author}{Sagawa, T.},
  \bibinfo{author}{Ueda, M.}, \bibinfo{author}{Muneyuki, E.},
  \bibinfo{author}{Sano, M.}, \bibinfo{year}{2010}.
\newblock \bibinfo{title}{Experimental demonstration of information-to-energy
  conversion and validation of the generalized {J}arzynski equality}.
\newblock \bibinfo{journal}{Nature Physics} \bibinfo{volume}{6},
  \bibinfo{pages}{988--992}.
\bibitem[{Wheeler(1995)}]{wheeler}
\bibinfo{author}{Wheeler, J.A.}, \bibinfo{year}{1995}.
\newblock \bibinfo{title}{Toward ``it from bit''}, in:
  \bibinfo{editor}{Anandan, J.S.}, \bibinfo{editor}{Safko, J.L.} (Eds.),
  \bibinfo{booktitle}{Quantum Coherence and Reality}, \bibinfo{publisher}{World
  Scientific}. p. \bibinfo{pages}{281}.
\newblock \DOIprefix\doi{10.1142/9789814533294}.
\bibitem[{Wolpert(2019)}]{wolpert2019}
\bibinfo{author}{Wolpert, D.H.}, \bibinfo{year}{2019}.
\newblock \bibinfo{title}{The stochastic thermodynamics of computation}.
\newblock \bibinfo{journal}{Journal of Physics A: Mathematical and Theoretical}
  \bibinfo{volume}{52}, \bibinfo{pages}{193001}.
\newblock \DOIprefix\doi{10.1088/1751-8121/ab0850}. \bibinfo{note}{publisher:
  IOP Publishing}.
\bibitem[{Zurek(1990)}]{Zurek1984}
\bibinfo{author}{Zurek, W.}, \bibinfo{year}{1990}.
\newblock \bibinfo{title}{Maxwell's demon, {S}zilard's engine and quantum
  measurements}, in: \bibinfo{editor}{Leff, H.S.}, \bibinfo{editor}{Rex, A.F.}
  (Eds.), \bibinfo{booktitle}{Maxwell's Demon: Entropy, Information,
  Computing}. \bibinfo{publisher}{Princeton University Press, New Jersey}.
\bibitem[{Zurek(2003)}]{Zurek2003}
\bibinfo{author}{Zurek, W.}, \bibinfo{year}{2003}.
\newblock \bibinfo{title}{Quantum discord and {M}axwell's demons}.
\newblock \bibinfo{journal}{Phys. Rev. A} \bibinfo{volume}{67},
  \bibinfo{pages}{012320}.

\end{thebibliography}



\end{document}